\documentclass{article}

\usepackage{amsthm}
\usepackage{graphicx}
\usepackage{float}
\usepackage{color}
\usepackage{amsmath}
\usepackage{amssymb}
\usepackage{multirow}
\usepackage{multicol}
\usepackage{authblk}
\usepackage{array}
\usepackage{epstopdf}

\author[1]{Glenn Young}
\author[1]{Andrew Belmonte}
\affil[1]{Department of Mathematics, Pennsylvania State University, University Park, PA 16802}
\date{}                     %% if you don't need date to appear
\begin{document}

\title{Fast cheater migration stabilizes coexistence \\
in a public goods dilemma on networks}
\date{\today}

\maketitle
%\maketitle

\section*{Abstract}
%A public good is any commodity that is available to the the general public that require some personal cost to produce.  Evolutionary game theory predicts that those who choose not to contribute to the public good, typically called ``cheaters,'' will exploit those who do, called ``cooperators,'' in a society that depends on public goods to survive, thereby causing the collapse of cooperation and, by extension, of that society. Organisms that can maintain cooperation despite facing a public goods dilemma are therefore of much interest to evolutionary ecologists. The ant species \emph{Pristomyrmex punctatus} is an example of such an organism. Workers of this species reproduce, care for broods, and forage outside of the nest, while genetic cheaters only reproduce. One proposed mechanism by which these ants might avoid colony collapse is their ability to migrate. Cheaters ants tend to migrate at a much faster rate than workers, an advantage that could allow them to avoid their own extinction by emigrating from a failing nest. 

Through the lens of game theory, cooperation is frequently considered an unsustainable strategy: if an entire population is cooperating, each individual can increase its overall fitness by choosing not to cooperate, thereby still receiving all the benefit of its cooperating neighbors while no longer expending its own energy. Observable cooperation in naturally-occurring public goods games is consequently of great interest, as such systems offer insight into both the emergence and sustainability of cooperation. Here we consider a population that obeys a public goods game on a network of discrete regions (that we call nests), between any two of which individuals are free to migrate. We construct a system of piecewise-smooth ordinary differential equations that couple the within-nest population dynamics and the between-nest migratory dynamics. Through a combination of analytical and numerical methods, we show that if the workers within the population migrate sufficiently fast relative to the cheaters, the network loses stability first through a Hopf bifurcation, then a torus bifurcation, after which one or more nests collapse. Our results indicate that fast moving cheaters can act to stabilize worker-cheater coexistence within  network that would otherwise collapse. We end with a comparison of our results with the dynamics observed in colonies of the ant species \emph{Pristomyrmex punctatus} and in those of the Cape honeybee {\it Apis mellifera capensis}, and argue that they qualitatively agree.

\section*{Abstract}

%%%%%%%%%%%%%%%%%%%%%%%%%%%%%%%%%%%%%%%%%%%%%%%%%%%%%%%%%
\section{Introduction}
%
%Paragraphs on
%\begin{itemize}
%\item Ecological motility\checkmark
%\item \emph{Pristomyrmex punctatus}, Dobata\checkmark
%\end{itemize}

Public goods dilemmas occur when an individual must choose whether or not to contribute to a commonly available (public) good \cite{Archetti2,Fehr,Hauert,Oakland,Perc,Wakano}. Contributing benefits the population by increasing the total amount of the public good, but it comes at a cost to the individual (through energy expenditure, for example), whereas choosing not to contribute comes at no cost to the individual, who still shares in benefits from the public good. Thus individuals must choose between what is best for themselves and what is best for the population, and in the basic public goods game, benefitting without contributing is strictly speaking always the better choice \cite{Archetti2,Perc}. 
Consequently, the dilemma inherent in public goods games have been used as a framework for studying the origin of cooperation and other group interactions \cite{Archetti2,levin2014}.  Such dilemmas are most naturally found in biological systems involving populations living closely enough together so that individual efforts are inevitably shared, for instance in bacterial species \cite{Damore} such as \emph{Escherichia coli} \cite{Vulic} and \emph{Myxococcus xanthus} \cite{Velicer}, in social insects such as ants \cite{Dobata1} and bees \cite{Martin}, and even in tumor cells, in which a subpopulation of cells produce an insulin-like tumor growth factor \cite{Archetti1,Gerlee}. While some such systems manage to persist in the presence of parasitic cheaters (e.g., the queenless ant species \emph{Pristomyrmex punctatus}, \cite{Dobata1}), others collapse (e.g., the Cape honey bee \cite{Martin}), leading to the natural question of what mechanisms foster these two outcomes. Many such mechanisms for sustaining cooperation have been studied, most notably rewarding cooperation \cite{Szolnoki} and punishing defection \cite{Fehr,Riehl}, or through interspecies competition: competition with a common opponent can stabilize cooperation within a public goods game within microbial species \cite{Celiker}. None of these have been observed in either of the previously mentioned ant or bee systems \cite{Dobata2,Martin}, indicating that there are still unexplored mechanisms by which cooperation can be sustained.

Our focus is on species that reproduce via parthenogenesis and occupy two or more discrete regions, such as nests or colonies, between which individuals are free to migrate.  Parthenogenesis is a form of asexual reproduction in which an embryo develops into an organism without fertilization \cite{mittwoch1978}. The inheritance of behavioral phenotypes this makes possible allows for a simplification in modeling reproductive competition mathematically, since differences in fitness impact survival directly. Moreover, parthenogenetic reproduction is common in nature, and has been observed in species ranging from aphids \cite{simon2002}, ants \cite{Dobata1,Dobata2,Tsuji}, and bees \cite{Martin} to zebra sharks \cite{dudgeon2017} and Komodo dragons \cite{watts2006}.

Migration often plays a large role in many species' survival. Driven by internal or external stimuli, species as simple as bacteria \cite{Turnbull} to complex organisms such as fish \cite{Partridge}, birds \cite{Thompson}, and countless others depend on their collective ability to migrate to evade environmental or ecological stress. Mathematical models for migration have expanded our understanding of the formation of spatial patterns in bacteria \cite{Tyson}, flocking behavior of birds \cite{Heppner}, and nest-site selection in honeybees \cite{Reina}. Here, we connect the well-studied fields of public goods games with such models of migratory dynamics to study migration as a mechanism to stabilize cooperation.

%\imp{Colonies of the queenless ant species \emph{Pristomyrmex punctatus} in Japan maintain a subpopulation of genetic ``cheaters'' that contribute nothing to the public good, in contrast with the productive ``workers,'' who forage outside of the nest for food and tend to the young within the nest \cite{Dobata1,Dobata2}. The cheaters naturally add stress to the workers in the form of increased demand on these public efforts. Not surprisingly, these cheaters spread between colonies, taking advantage of the free public goods while harming the overall fitness of each colony \cite{Dobata1,Tsuji}. This characterization stems in part from the estimation that cheaters tend to migrate between nests at a rate roughly three orders of magnitude faster than their worker counterparts, suggesting that cheaters can abandon a nest before it collapses due to their own presence \cite{Dobata1,Dobata2,Tsuji}. }

For the remainder of this paper, we will call the individuals that cooperate ``workers,'' those that do not ``cheaters,'' and the discrete regions they occupy ``colonies'' or ``nests.''   To model the role that cheaters play in such populations, we propose a mathematical model coupling within-colony population dynamics and between-colony migratory dynamics. While our model shows that cheaters drive down the fitness of an individual colony, it also makes the unexpected prediction that fast cheaters can actually \emph{stabilize} a network of colonies by saving individual colonies from collapse.
The remainder of this paper is organized as follows: in the following section, we develop a system of ordinary differential equations modeling the population dynamics of workers and cheaters in a network of $N$ colonies, connected according to a given connectivity matrix $B$. We then analyze the behavior of a single colony, to determine conditions under which workers and cheaters can coexist. In Section \ref{sec:twoNest}, we begin our investigation of the effects of migration by considering a two-colony network, between which the worker and cheater populations are free to travel, and determine conditions under which the two nest colony can be maintained, and conditions under which it collapses into a single nest. Section \ref{sec:nNest} builds on the results and intuition of the two-colony system to determine similar conditions under which an 
$N$-colony, all-to-all connected network collapses. Finally, we consider $N$-colony networks connected according to more complicated graph structures, and discuss the implications of our results.

%\newpage
\section{Model construction}\label{sec:model}

In this section, we develop a model of within- and between-colony dynamics of a parthenogenetically reproducing organism. We generally consider a collection of $N$ colonies, which we will hereafter refer to as a colony network, or simply a \emph{network}, in which individual colonies are connected according to connectivity matrix $B=[\beta_{ij}]$: colony $i$ is directly connected to colony $j$ if and only if $\beta_{ij}\not=0$, and $\beta_{ij}>0$ defines a migration rate between the two colonies.  

Within each colony, the population is comprised of two sub-populations: \emph{workers}, who contribute to the public good, and \emph{cheaters}, who do not, the two of which interact according to a variation of an ecological public goods game \cite{Hauert,Wakano}. We denote by $u_i$ the population of workers and by $v_i$ population of cheaters in nest $i$. 
%We will first determine pertinent within-nest population dynamics, then add in between-nest migratory dynamics.
%
Within this nest, the worker and cheater populations are assumed to grow proportionate to their respective population sizes:
\begin{equation*}
\begin{aligned}
\dot u_i &=  u_iF_i(u_i,v_i)\\
\dot v_i &=  v_iG_i(u_i,v_i),
\end{aligned}
\end{equation*}
where $F$ and $G$ are their respective fitnesses. We assume each population grows according to the logistic model, and therefore

\begin{equation}\label{eq:log}
\begin{aligned}
F_i(u_i,v_i)&=f(u_i,v_i)-\gamma_i(u_i+v_i)-\mu_u\\
G_i(u_i,v_i)&=g(u_i,v_i)-\gamma_i(u_i+v_i)-\mu_v,
\end{aligned}
\end{equation}
where $f$ and $g$ denote the growth rates of $u_i$ and $v_i$, respectively, $\mu_u$, $\mu_v$ denote the death rates, and $\gamma_i$ defines the carrying capacity of colony $i$. In this way, $\gamma_i$ can be interpreted as the quality of the $i$th nest, with smaller $\gamma_i$ corresponds to a nest of higher quality.

We further assume that the growth of each population ($f$, $g$) is proportional to their respective payouts from a public goods game. If each worker contributes $c$ to the public good, then $cu_i/(u_i+v_i)$ is available for each individual within the nest on average \cite{Hauert}. We therefore define 
$$f(u_i,v_i)=\frac{r_ucu_i}{u_i+v_i}-c,$$ and $$g(u_i,v_i)=\frac{r_vcu_i}{u_i+v_i},$$ where $r_u$ and $r_v$ are the intrinsic growth rates of the workers and cheaters, respectively. Combining these with System (\ref{eq:log}) produces the following fitness functions:
\begin{equation*}
\begin{aligned}
F_i(u_i,v_i)&=\frac{r_ucu_i}{u_i+v_i}-c-\gamma_i(u_i+v_i)-\mu_u\\
G_i(u_i,v_i) &= \frac{r_vcu_i}{u_i+v_i}-\gamma_i(u_i+v_i)-\mu_v,
\end{aligned}
\end{equation*}
and therefore the within-nest dynamics are governed by the system
\begin{equation}\label{eq:withinNest}
\begin{aligned}
\dot{u}_i&=u_i\left[\frac{r_ucu_i}{u_i+v_i}-c-\gamma_i(u_i+v_i)-\mu_u\right]\\
\dot{v}_i &= v_i\left[\frac{r_vcu_i}{u_i+v_i}-\gamma_i(u_i+v_i)-\mu_v\right],
\end{aligned}
\end{equation}
which is similar to the ecological public goods models found in \cite{Hauert,Wakano}.

A shortcoming of this model is that it is undefined at the extinction state $(u_i^*,v_i^*)=(0,0)$, although bounded \cite{deForest2}. Because we are interested in the possibility of extinction, we remedy this by redefining the growth functions $f$ and $g$ within the fitness functions $F_i$ and $G_i$, respectively, as $$f(u_i,v_i)=\frac{r_ucu_i}{u_i+v_i+\epsilon}-c,$$ and $$g(u_i,v_i)=\frac{r_vcu_i}{u_i+v_i+\epsilon},$$ where $0<\epsilon<<1$. The parameter $\epsilon$ can be thought of as a decay rate in the public good, see Appendix \ref{sec:epsilon}. With this modification, the system for the within-nest-$i$ dynamics becomes

\begin{equation}\label{eq:noMigEps}
\begin{aligned}
\dot u_i &= u_i\left[\frac{r_ucu_i}{u_i+v_i+\epsilon}-c-\gamma_i(u_i+v_i)-\mu_u\right]\\
\dot v_i &= v_i\left[\frac{r_vcu_i}{u_i+v_i+\epsilon}-\gamma_i(u_i+v_i)-\mu_v\right].
\end{aligned}
\end{equation}

We now turn our attention to between-nest migration. In their most general form, migratory dynamics can be included in our model as follows
\begin{equation}\label{eq:nNest}
\begin{aligned}
\dot u_i &=  u_iF_i(u_i,v_i)+\displaystyle\sum_{j=1}^N \alpha_{ij}H_u(u_i,v_i,u_j,v_j)\\
\dot v_i &=  v_iG_i(u_i,v_i)+\displaystyle\sum_{j=1}^N \beta_{ij}H_v(u_i,v_i,u_j,v_j),
\end{aligned}
\end{equation}
for $i=1\dots N$, where $\alpha_{ij}$ and $\beta_{ij}$ denote the migration rates between nests $i$ and $j$ of the workers and cheaters, respectively. The functions $H_u$ and $H_v$ define the migration dynamics of the worker and cheaters, respectively. We assume that individuals will prefer to move from nests of lower fitness to nests of higher fitness, and we therefore choose the $H$ functions so that individuals move in the direction that maximizes fitness gain \cite{Cantrell,cosner2005,Cressman,deForest1,deForest2,Shigesada}. Under this assumption, and following \cite{deForest2} we define
\begin{equation}\label{eq:migration}
\begin{aligned}
H_u(u_i,v_i,u_j,v_j)&=\left[F_i(u_i,v_i)-F_j(u_j,v_j)\right]\delta u_{ij} 
\quad \textrm{ and}\\
H_v(u_i,v_i,u_j,v_j)&=\left[G_i(u_i,v_i)-G_j(u_j,v_j)\right]\delta v_{ij},
\end{aligned}
\end{equation}
where 
\begin{equation*}
\delta u_{ij}=\begin{cases} 
      u_j, & \text{if } F_i(u_i,v_i)-F_j(u_j,v_j)\geq0 \\
      u_i, & \text{if } F_i(u_i,v_i)-F_j(u_j,v_j)<0
   \end{cases}
\end{equation*}
and
\begin{equation*}
\delta v_{ij}=\begin{cases} 
      v_j, & \text{if } G_i(u_i,v_i)-G_j(u_j,v_j)\geq0 \\
      v_i, & \text{if } G_i(u_i,v_i)-G_j(u_j,v_j)<0.
   \end{cases}
\end{equation*}
The terms $\delta u_{ij}$ and $\delta v_{ij}$ guarantee that the migratory flux is proportional to the population size within the nest of lower fitness; that is, the nest from which individuals are emigrating \cite{deForest2}. Consequently, the system is nonsmooth along the manifolds defined by $F_i(u_i,v_i)-F_j(u_j,v_j)=0$ and $G_i(u_i,v_i)-G_j(u_j,v_j)=0$ (commonly referred to as ``switching manifolds,'' see, e.g., \cite{Harris,Leine}) for each pair $i\not=j$. Though this lack of smoothness complicates the analysis in Sections \ref{sec:twoNest} and \ref{sec:nNest}, we are still able to extract important information concerning the stability of a colony in specific cases, and use this information to analyze general cases.

%%%%%%%%%%%%%%%%%%%%%%%%%%%%%%%%%%%%%%%%%%
\subsection{Single colony population dynamics}\label{sec:1nest}

In order to study the effects migratory dynamics have on networks of nests, we must first understand the behavior of a single nest without any such dynamics, 
%We therefore remove the  migratory dynamics from system (\ref{eq:nNest}), which reduces to 
System (\ref{eq:noMigEps}). Since we are considering a single nest, we suppress the subscripted $i$'s in this section and denote the worker and cheater populations by $u$ and $v$, respectively. 

Note that this system does not have the nonsmooth manifolds mentioned above, and consequently standard analytical methods for continuous dynamical systems can be applied. Before we consider this system in its entirety, we first  examine the system in the absence of cheaters, to establish a condition that guarantees the survival of the workers in a cheater-free nest:

\begin{equation}\label{eq:noCheaters}
\dot u=u\left[r_uc-c-\gamma u-\mu_u\right].
\end{equation}
Equation (\ref{eq:noCheaters}) is the standard logistic growth model with effective growth rate $r_u c-(c+\mu_u)$. The worker population will tend toward its carrying capacity $[r_uc-(c+\mu_u)]/\gamma$ as long as $r_u c-(c+\mu_u)>0$, and will otherwise tend toward extinction. We will consequently impose the condition $r_u c-(c+\mu_u)>0$ in Systems (\ref{eq:noMigEps}) and (\ref{eq:nNest}).

We now include the presence of cheaters in System (\ref{eq:noMigEps}). In the  classic public goods game, each individual is assumed to derive the same payout from the public good; that is $r_u=r_v$ \cite{Archetti2,Hauert,Wakano}. Under this assumption, the asymptotic behavior of the two populations is entirely determined by the difference in their respective effective decay rates, specifically by the sign of $\mu_u+c-\mu_v$. If this quantity is negative, the workers will outcompete the cheaters. If the quantity is negative, the cheaters outcompete the workers, but because the workers are the only population contributing to the public good, both populations necessarily tend toward extinction. Biologically, the parameters $r_u$ and $r_v$ can be thought of as the relative growth rate of populations $u$ and $v$, respectively, when each worker is contributing to the public good at rate $c$. It is reasonable, then, to assume that $r_u\not= r_v$ in general, as workers and cheaters are often phenotypically distinct and therefore have potentially different rates of growth \cite{Dobata2,Tsuji}. Without loss of generality, and for consistency with \cite{Dobata2,Tsuji}, we will assume that $r_u<r_v$. %, based on reported growth rates of \emph{Pristomyrmex punctatus} \cite{Dobata2}. 
Reversing this inequality does not change any of the qualitative behaviors described below, only the regions in parameter space in which they are observed.

%%% THIS PART SOONER
 
%System (\ref{eq:noMig}) only supports a single equilibrium point in the nonnegative quadrant, namely
% % $$(u,v)=\left(\frac{r_uc-(c+\mu_u)}{\gamma},0\right)$$ and  
% $$(u,v)=\frac{r_u\mu_v-r_v(\mu_u+c)}{c\gamma(r_v-r_u)^2}\left(\mu_v-(\mu_u+c),c(r_v-r_u)-(\mu_v-(\mu_u+c))\right).$$
%An important outcome that our model must be able to capture is extinction of one or both populations. However, the system is singular at the extinction equilibrium $(u_0^*,v_0^*)=(0,0)$, and consequently an initially positive population within a nest cannot become extinct. To remedy this, we redefine the growth functions $f$ and $g$ as $$f(u_i,v_i)=\frac{r_ucu_i}{u_i+v_i+\epsilon}-c,$$ and $$g(u_i,v_i)=\frac{r_vcu_i}{u_i+v_i+\epsilon},$$ where $0<\epsilon<<1$. The parameter $\epsilon$ can be thought of as a decay rate in the public good, and we assume it is small (see Appendix \ref{sec:epsilon}). With this modification, the system we study becomes
%
%\begin{equation}\label{eq:noMigEps}
%\begin{aligned}
%\dot u_i &= u_i\left[\frac{r_ucu_i}{u_i+v_i+\epsilon}-c-\gamma_i(u_i+v_i)-\mu_u\right]\\
%\dot v_i &= v_i\left[\frac{r_vcu_i}{u_i+v_i+\epsilon}-\gamma_i(u_i+v_i)-\mu_v\right].
%\end{aligned}
%\end{equation}
System (\ref{eq:noMigEps}) supports four equilibria in the nonnegative quadrant:  extinction  
\begin{equation*}\label{eq:E0}\tag{E$_0$}
\begin{aligned}
u_0^*&=0\\
v_0^*&=0,
\end{aligned}
\end{equation*} two equilibria which have only workers 
\begin{equation*}\label{eq:E1}\tag{E$_1$}
\begin{aligned}
u_1^*&=\frac{cr_u-(\mu_u+c)-\epsilon\gamma-\sqrt{(cr_u-(\mu_u+c)-\epsilon\gamma)^2-4\epsilon\gamma(\mu_u+c)}}{2\gamma}\\
v_1^*&=0,
\end{aligned}
\end{equation*} and
\begin{equation*}\label{eq:E2}\tag{E$_2$}
\begin{aligned}
u_2^*&=\frac{cr_u-(\mu_u+c)-\epsilon\gamma+\sqrt{(cr_u-(\mu_u+c)-\epsilon\gamma)^2-4\epsilon\gamma(\mu_u+c)}}{2\gamma}\\
v_2^*&=0,
\end{aligned}
\end{equation*}
and the coexistence equilibrium
\begin{equation*}\label{eq:E3}\tag{E$_3$}
\begin{aligned}
u_3^*&=\frac{(\mu_v-(\mu_u+c))(r_u\mu_v-r_v(\mu_u+c)+\epsilon\gamma(r_v-r_u))}{c\gamma(r_v-r_u)^2}\\
v_3^*&=\frac{(r_u\mu_v-r_v(\mu_u+c))(c(r_v-r_u)+\mu_u+c-\mu_v)}{c\gamma(r_v-r_u)^2}-\epsilon\frac{\mu_v-(\mu_u+c)}{c(r_v-r_u)}.
\end{aligned}
\end{equation*}
For $\epsilon>0$ small, equilibria E$_1$ and E$_2$ can be approximated by
\begin{equation*}\label{eq:approxE1}
(u_1^*,v_1^*)=\left(\frac{\mu_u+c}{cr_u-(\mu_u+c)}\epsilon,0\right),%\tag{$\sim$E$_1$}
\end{equation*}
and
\begin{equation*}\label{eq:approxE2}
(u_2^*,v_2^*)=\left(\frac{cr_u-(\mu+c)}{\gamma}-\frac{cr_u}{cr_u-(\mu_u+c)}\epsilon,0\right).%\tag{$\sim$ E$_2$}
\end{equation*}
These approximations simplify the stability calculations below, and make clear that E$_1$ is near the origin. Importantly, equilibrium point E$_0$ is not singular in system (\ref{eq:noMigEps}), and extinction is consequently both well defined and possible.

Standard linear stability analysis reveals that, as long as all parameters are positive, the extinction equilibrium E$_0$ is always stable, and therefore the nest can always collapse if the initial worker population is sufficiently small. Similarly, E$_1$ is always unstable, either as a node or saddle point. The coexisting equilibrium point E$_3$ is stable if and only if 
\begin{equation}\label{eq:E3stab}
r_u\mu_v>r_v(\mu_u+c) \textrm{ and } c(r_v-r_u)+(\mu_u+c-\mu_v)>0.
\end{equation} 
Interestingly, the stability of E$_3$ does not depend on the nest-quality parameter $\gamma$. This means that no matter how poor the quality of the nest might be, the worker and cheater populations can coexist there, as long as condition (\ref{eq:E3stab}) is satisfied. Therefore, any nest collapse observed once migratory dynamics are introduced will be entirely a consequence of the migration itself (see Sections \ref{sec:twoNest} and \ref{sec:nNest}). We will generally use the parameter values found in Table \ref{tab:params}, which satisfy (\ref{eq:E3stab}), unless otherwise specified.

These steady states provide the first insight into the effect cheaters have on a nest by allowing us to compare the total population before and after cheater invasion. For simplicity, we consider the case when $\epsilon=0$. The ratio $\rho$ of the total population size at steady state after invasion to the total population size at steady state before invasion is
\begin{equation*}
\begin{aligned}
\rho&=\frac{u_3^*+v_3^*}{u_1^*}
&=\frac{r_u\mu_v-r_v(\mu_u+c)}{(r_v-r_u)(cr_u-(\mu_u+c))}.
\end{aligned}
\end{equation*}
Under stability condition (\ref{eq:E3stab}), $0<\rho<1$, and perhaps unsurprisingly, cheaters necessarily harm the population by reducing its total sustainable size within the nest. For the parameters in Table \ref{tab:params}, $\rho=0.25$, which means that the presence of cheaters reduces the total population size by $75\%$ in this case.

\begin{table}[H]
{\centering
\begin{tabular}{ p{.25in}  p{2.5in}  p{.75in} }
\hline
\multicolumn{2}{c}{Parameter} & Default value\\
\hline\hline \noalign{\smallskip}
$r_u$ & Growth rate of workers & 5\\
$r_v$ & Growth rate of cheaters & 6\\
$c$ & Diffusivity of nutrient & 1\\
$\gamma_i$ & Quality of nest $i$ & 1\\
$\mu_u$ & Mortality rate of workers & 2\\
$\mu_v$ & Mortality rate of cheaters & 3.7\\
$\beta_u$ & Migration rate of workers & varies\\
$\beta_v$ & Migration rate of cheaters & varies\\
\end{tabular}
\caption{Parameters used in model (\ref{eq:nNest}) unless otherwise indicated.}
\label{tab:params}

}
\end{table}

Stability condition (\ref{eq:E3stab}) can be interpreted as a ``happy medium'' of cheater fitness when contrasted with the alternative conditions in Table \ref{tab:stab}.  Rewriting the conditions slightly to ease biological interpretation, the worker-only equilibrium E$_2$ is stable if and only if $r_u/(\mu_u+c)>r_v/\mu_v$ and $cr_u-(\mu_u+c)>cr_v-\mu_v$. In this case, the worker population is much more fit than the cheater population in terms of relative growth rate. Likewise, if $r_u/(\mu_u+c)<r_v/\mu_v$ and $cr_u+(\mu_u+c)<cr_v-\mu_v$, the cheater population is much more fit than the worker population. In this case, the cheaters outcompete the workers, and consequently collapse the entire nest; that is, the extinction state E$_0$ is the only stable equilibrium. 

%\begin{table}[H]
%\begin{center}
%\begin{tabular}{| l | l | l |}
%\hline
% & $c(r_v-r_u)+(\mu_u+c-\mu_v)>0$ & $c(r_v-r_u)+(\mu_u+c-\mu_v)<0$ \\ 
%\hline
% & $(u_1,v_1)$ saddle & $(u_1,v_1)$ saddle \\ 
%$r_u\mu_v>r_v(\mu_u+c)$ & $(u_2,v_2)$ saddle & {\bf $(u_2,v_2)$ stable}\\
% &  {\bf $(u_3,v_3)$ stable} & $(u_1,v_1)$ saddle, $v_3<0$\\
%\hline
% & $(u_1,v_1)$ unstable node &  \\ 
%$r_u\mu_v<r_v(\mu_u+c)$ & $(u_2,v_2)$ saddle & Impossible\\
% &   $(u_3,v_3)$ saddle, $v_3<0$ & \\
% \hline
%\end{tabular}
%\end{center}
%\caption{Stability behavior of equilibria.}
%\label{tab:stab}
%\end{table}
%Note that the system is bistable as long as $r_u\mu_v>r_v(\mu_u+c)$. In this case, the equilibrium $(u_1,v_1)$ is a saddle point, the stable manifold of which acts as a separatrix between extinction and persistence (Fig \ref{fig:stabMan}). 

\begin{figure}[H]
{\centering
\includegraphics[width=2.45in]{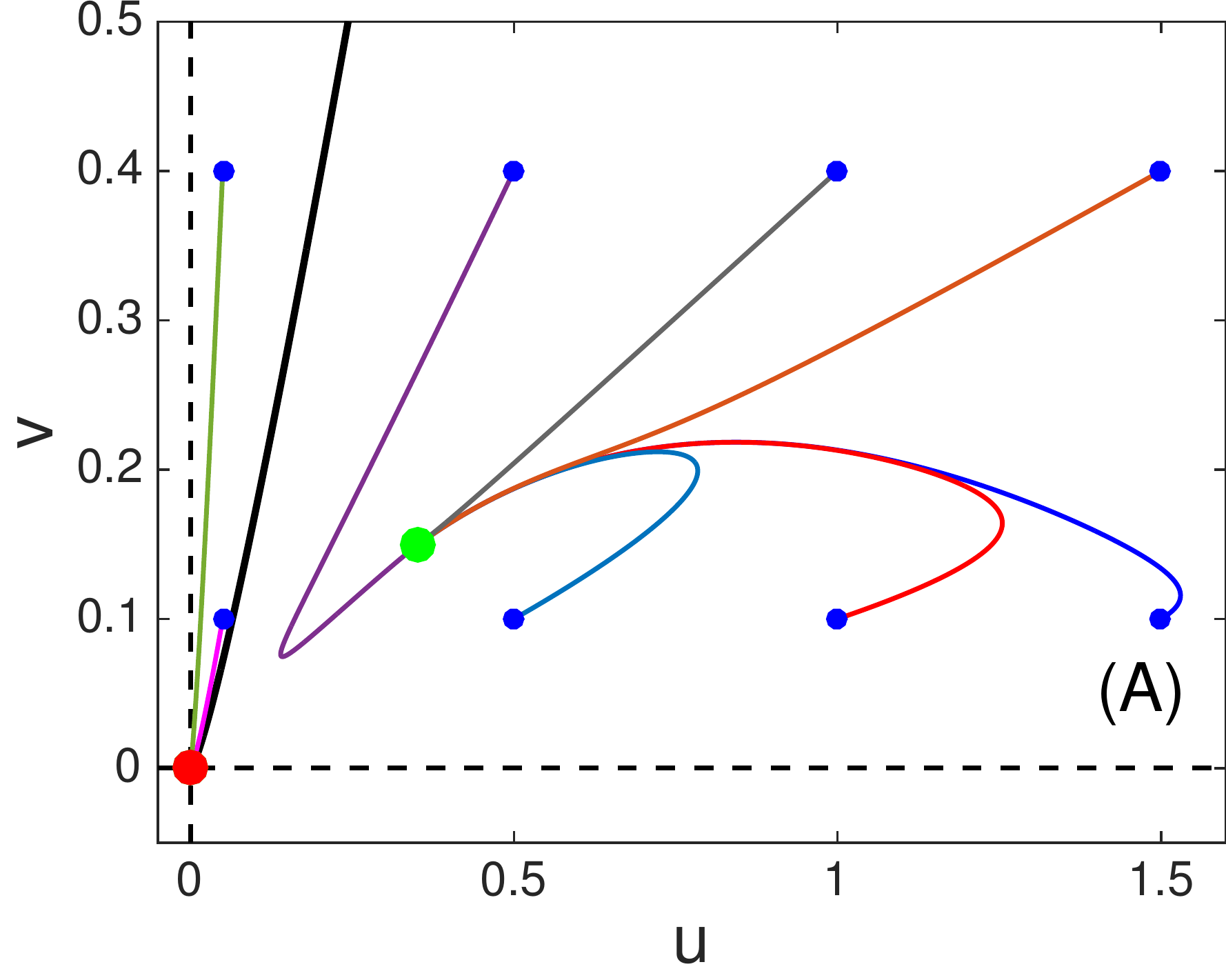} %\includegraphics[width=2.3in]{oneNestvsTime.eps}   
\includegraphics[width=2.4in]{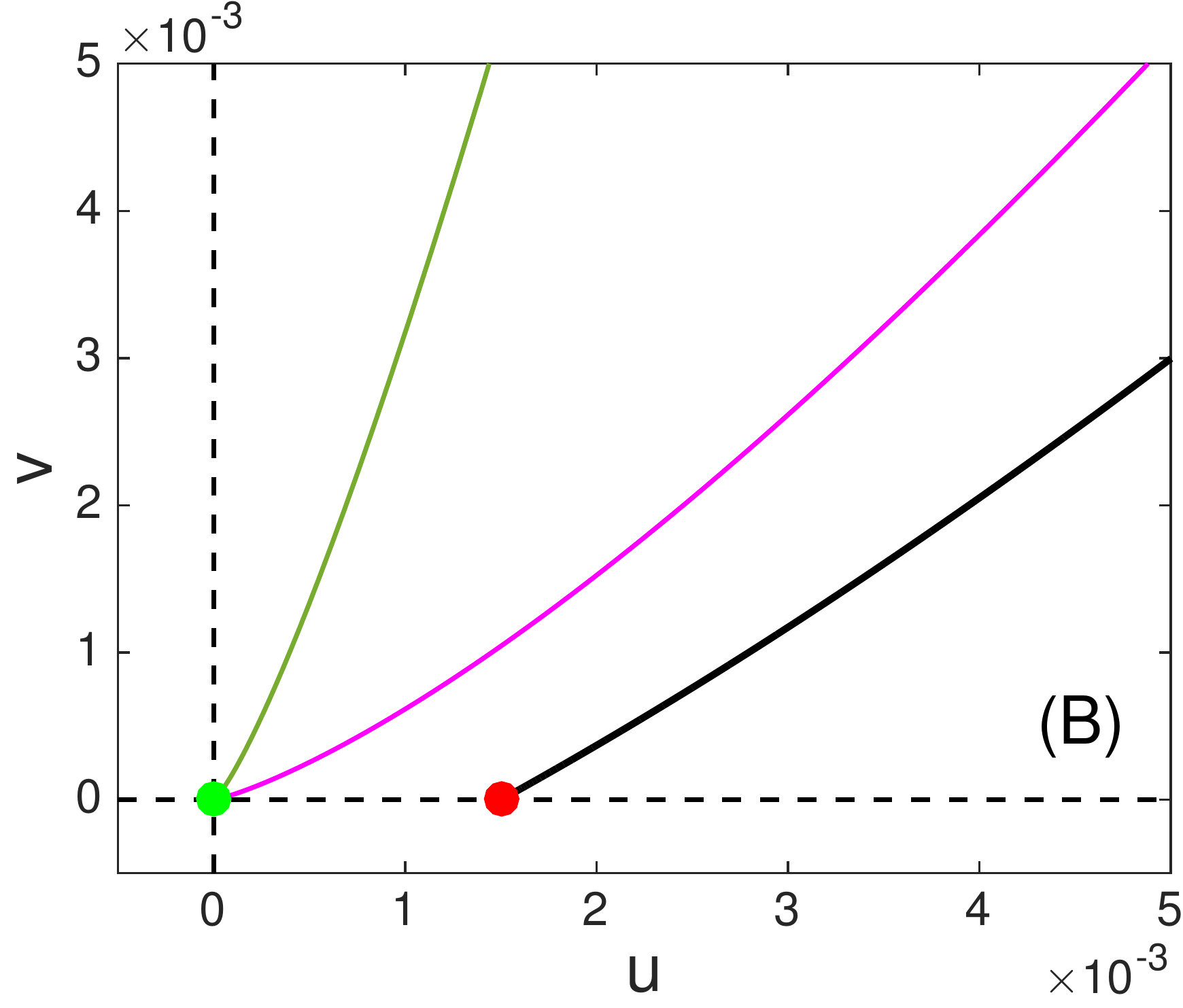} 
}
\caption{Within-nest dynamics, for parameters listed in Table \ref{tab:params}: A. Sample trajectories in the $uv$-plane. The red and green circles denote equilibrium points E$_1$ and E$_3$, respectively, and the blue circles represent the starting position of each solution. The thick black curve is the stable manifold of E$_1$. To the left of the curve both the worker and cheater populations eventually become extinct, to the right, both populations approach E$_3$ and consequently persist; B. Equilibria E$_0$ and E$_1$. Zooming in on phase plane near the origin shows the stable extinction state E$_1$ (green circle) and the saddle point E$_1$ (red circle). The stable manifold emanating from E$_1$ is acts as the separatrix between the two stable equilibria E$_1$ and E$_3$.}%Sample trajectories over time. The color of the curve corresponds to the solution of the same color in A. Solid lines represent the worker population $u$; dashed lines represent the cheater population $v$.}
\label{fig:stabMan}
\end{figure}

\begin{table}[H]
\begin{center}
\begin{tabular}{| l | l | l |}
\hline
 & $c(r_v-r_u)+(\mu_u+c-\mu_v)>0$ & $c(r_v-r_u)+(\mu_u+c-\mu_v)<0$ \\ 
\hline
 & E$_1$ saddle & E$_1$ saddle \\ 
$r_u\mu_v>r_v(\mu_u+c)$ & E$_2$ saddle & {\bf E$_2$ stable}\\
 &  {\bf E$_3$ stable} & E$_3$ saddle, $v_3<0$\\
\hline
 & E$_1$ unstable node &  \\ 
$r_u\mu_v<r_v(\mu_u+c)$ & E$_2$ saddle & Impossible\\
 &   E$_3$ saddle, $v_3<0$ & \\
 \hline
\end{tabular}
\end{center}
\caption{Stability behavior of equilibria.}
\label{tab:stab}
\end{table}
Because the extinction state E$_0$ is always stable, the results summarized in Table \ref{tab:stab} indicate that system (\ref{eq:noMigEps}) is bistable as long as $r_u\mu_v>r_v(\mu_u+c)$. In this case, equilibrium E$_1$ is a saddle point, the stable manifold of which acts as a separatrix between extinction and persistence (Figure \ref{fig:stabMan}). If in addition $c(r_v-r_u)+(\mu_u+c-\mu_v)>0$ (that is, stability condition (\ref{eq:E3stab}) is satisfied), then the coexistence state E$_3$ is stable. The separatrix is shown as the black curve emanating from the E$_1$ in Figure \ref{fig:stabMan}B. To the left of this curve, all solutions tend to the extinction equilibrium; that is, if the worker population is sufficiently small relative to the cheater population, the nest will collapse. To the right, they tend toward the coexistence steady state (Figure \ref{fig:stabMan}A). Nests will therefore only collapse if the cheater population becomes sufficiently large relative to the worker population. %Figure \ref{fig:stabMan}B shows a subset of the solution curves given in Figure \ref{fig:stabMan}A over time.

\section{Migration between two colonies}\label{sec:twoNest}
We begin our investigation of the effects of migration by considering a network comprised of only two colonies, between which all individuals are free to migrate. We assume that the migration rate from nest 1 to nest 2 is the same as the migration rate from nest 2 to nest 1 for both populations; that is, $\alpha_{12}=\alpha_{21}=\beta_u$ and $\beta_{12}=\beta_{21}=\beta_v$. System (\ref{eq:nNest}) therefore takes the form
\begin{equation}\label{eq:twoNest1}
\begin{aligned}
\dot u_1 &= \beta_u  H_u(u_1,v_1,u_2,v_2)+u_1F_1(u_1,v_1)\\
\dot v_1 &= \beta_v  H_v(u_1,v_1,u_2,v_2)+v_1G_1(u_1,v_1)\\
\dot u_2 &= -\beta_u  H_u(u_1,v_1,u_2,v_2)+u_2F_2(u_2,v_2)\\
\dot v_2 &= -\beta_v  H_v(u_1,v_1,u_2,v_2)+v_2G_2(u_2,v_2),
\end{aligned}
\end{equation}
where the $H$, $F$, and $G$ functions are as in system (\ref{eq:nNest}).

Both of the individual nests maintain the same four equilibria as in the one-nest system (\ref{eq:noMigEps}). This implies that system (\ref{eq:twoNest1}) has at least sixteen equilibrium points, though most are of little interest. Of particular importance are the extinction state $T_0=(0,0,0,0)$, the two states in which one nest collapses, $T_1=(u_1^*,v_1^*,0,0)$ and $T_2=(0,0,u_2^*,v_2^*)$, and the state $T^*=(u_1^*,v_1^*,u_2^*,v_2^*)$.  Linear stability analysis verifies that the extinction state $T_0$ is always stable and that under condition (\ref{eq:E3stab}), both $T_1$ and $T_2$ are also always stable. The system is therefore generally multistable. Because a single surviving nest necessarily tends to the within-nest coexisting state E$_3$, emphasis will now be shifted to nest survival or collapse, instead of the composition of populations within each nest. For the rest of this paper, we will therefore use the term \emph{coexisting state} to refer to $T^*$ above, or the $N$-nest equivalent state.

\begin{figure}[!b] % H !b
{\centering
\includegraphics[width=5.5in]{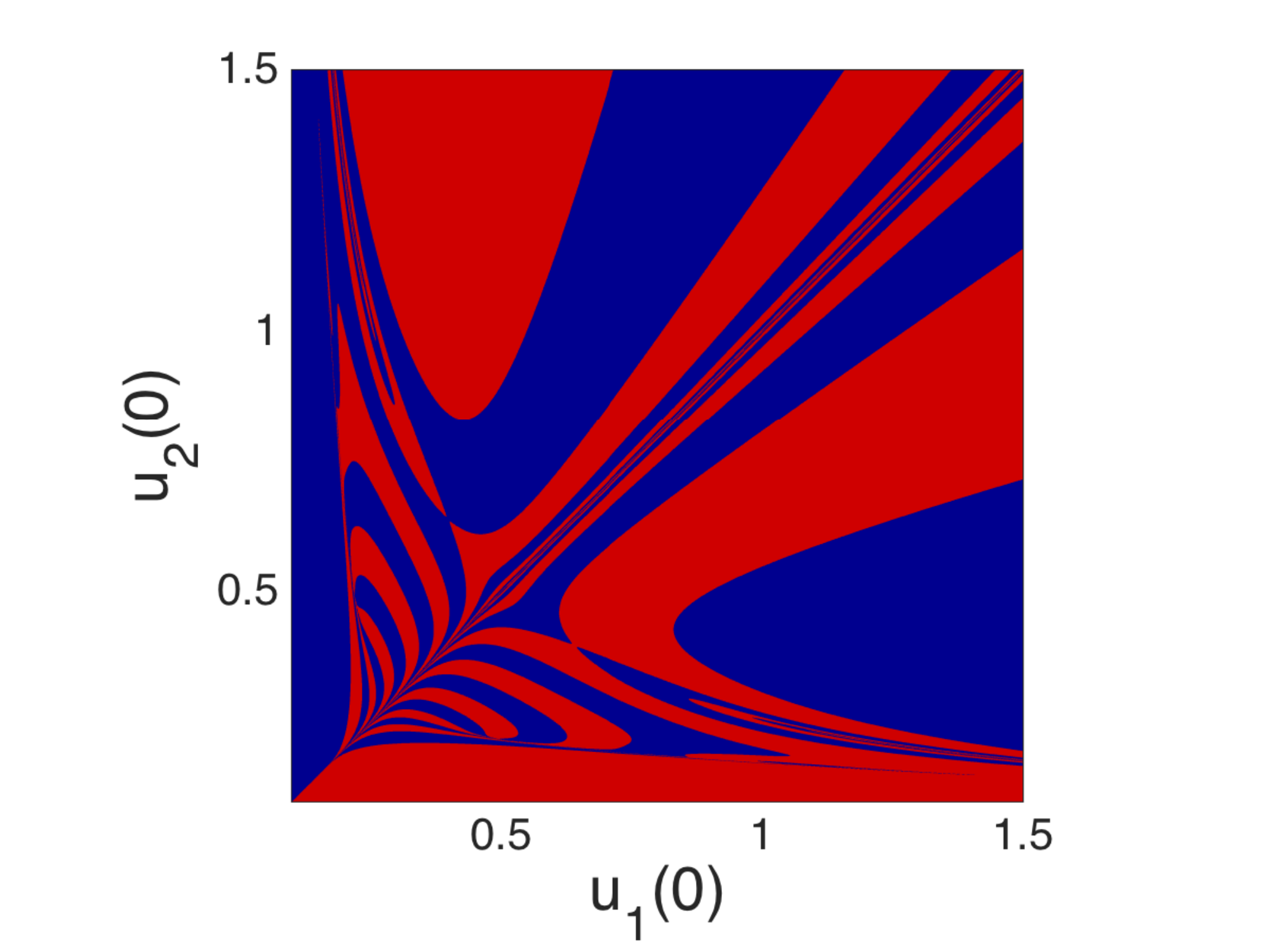}
}
\vspace{-4mm}
\caption{Basins of attraction for single nest final states, as a function of initial worker populations for two identical nests. The red region indicates the basin of attraction of the nest 1-only coexisting steady state $T_1$, and the blue region is the basin of attraction of the nest 2-only coexisting steady state $T_2$. The initial conditions for cheaters are fixed at $v_1(0)=v_2(0)=0.1$.}% \imp{this is a cool figure but doesn't really fit into the story very well...}}
\label{fig:basinSym}
\end{figure}

We now seek to determine the effect migration has on nest persistence and extinction. In general, both, one, or neither of the nests will collapse. In the absence of migration ($\beta_u=\beta_v=0$), the outcome of the two nests are independent of one another, and the longterm behavior is determined by the analysis presented in Section \ref{sec:1nest}. With nonzero migration, even in the simple case in which both nests are of equal quality, the system can show complicated dependence on parameters. For example, Figure \ref{fig:basinSym} shows the relationship between initial conditions and nest collapse for fixed migration rates between identical nests. 
%We therefore begin our analysis with the symmetric case in which both nests have the same carrying capacity.
Note that, unless $v_i(0) = 0$ in both nests, there will be cheaters in both nests for all $t$, so long as the nest does not collapse.

%%%%%%%%%%%%%%%

\subsection{Colonies of equal quality}

\begin{figure}[H]
{\centering
\includegraphics[width=2.6in]{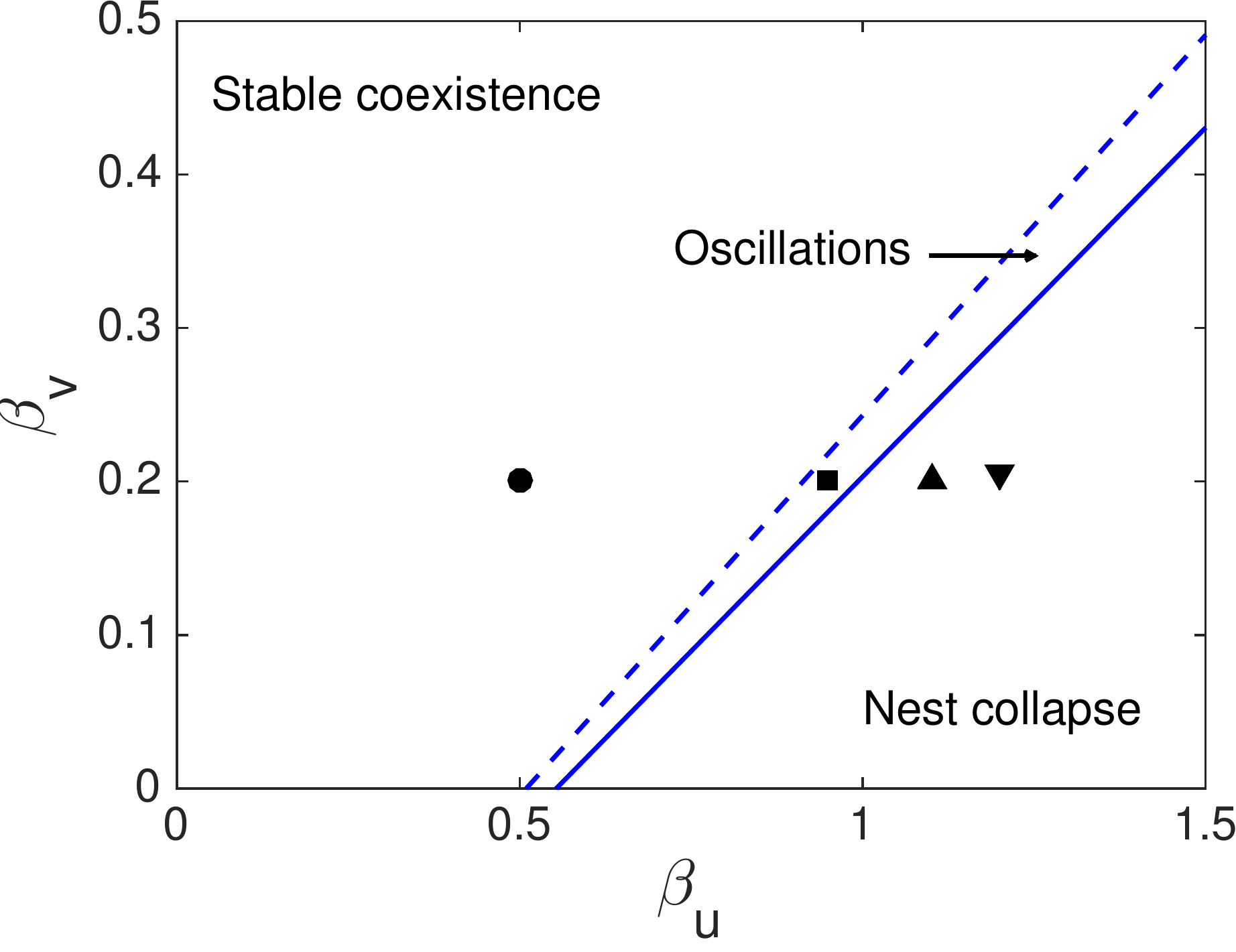}

}
\caption{Stability of the coexisting state $(u_1^*,v_1^*,u_2^*,v_2^*)$ over varied $\beta_u$ and $\beta_v$. The equilibrium is stable in region to the left of the dashed blue line. The dashed blue line corresponds to $\sigma=0$ and defines a line of supercritical Hopf bifurcations: for parameter values chosen between the dashed line and the solid line, trajectories are attracted to a stable periodic limit cycle.
% in the grey region. 
The limit cycle undergoes a torus bifurcation when $(\beta_u,\beta_v)$ passes through the solid blue line, destroying the stability of the limit cycle. In this region, exactly one of the two nests collapses. The points marked with shapes correspond to $(\beta_u,\beta_v)$ pairs considered in Figure \ref{fig:trajectories}.}
\label{fig:betaBif}
\end{figure}

In general, the interior equilibrium $T^*= (u_1^*,v_1^*,u_2^*,v_2^*)$ falls along the nonsmooth manifold, greatly complicating stability analysis \cite{Leine}.  However, in the nongeneric case in which the two nests are of the same quality ($\gamma_1=\gamma_2=\gamma$), we have $u_1^*=u_2^*=u^*$  and $v_1^*=v_2^*=v^*$, and the resulting symmetry about each nonsmooth manifold reduces the question of stability to that of standard linear stability analysis around this point.  The four-dimensional linearized system at this interior equilibrium has two complex pairs of eigenvalues: one pair is independent of $\beta_u$ and $\beta_v$ and has negative real part as long as (\ref{eq:E3stab}) is satisfied, and the other pair has real part
\begin{equation}\label{eq:re}
\sigma=\frac{1}{2}(1+2\beta_u)u^*\left(\frac{r_ucv^*}{(u^*+v^*)^2}-\gamma\right)-\frac{1}{2}(1+2\beta_v)v^*\left(\frac{r_vcu^*}{(u^*+v^*)^2}+\gamma\right).
\end{equation}
The interior equilibrium point undergoes a supercritical Hopf bifurcation when $\sigma$ changes from negative to positive. The left-most region (labeled ``Stable coexistence'') in Figure \ref{fig:betaBif}, corresponds to $(\beta_u,\beta_v)$ pairs such that $\sigma<0$ and $T^*$ is consequently asymptotically stable; that is, the two nests will generically maintain stable populations of coexisting workers and cheaters for motility rate pairs taken from this region. As $(\beta_u,\beta_v)$ values cross the dashed line, the equilibrium undergoes a Hopf bifurcation and solutions tend to a stable periodic limit cycle, in which workers and cheaters coexist in both nests for all time, but the  populations are constantly migrating back and forth between the nests. If $\beta_u$ is further increased, the limit cycle loses its stability through what is numerically determined to be a torus bifurcation.
The solid blue line in Figure \ref{fig:betaBif} denotes this line of torus bifurcations. For $(\beta_u,\beta_v)$ pairs chosen from the region to the right of the solid line, one nest necessarily collapses, leaving one nest in which workers and cheaters coexist (at equilibrium E$_3$, due to stability condition (\ref{eq:E3stab})). %\gy{We note here that if one nest collapses, the remaining nest will approach the coexisting steady state E$_3$ discussed in Section \ref{sec:1nest}, and will consequently never collapse. This is due to imposing stability condition (\ref{eq:E3stab}).}

\begin{figure}[H]
{\centering
\includegraphics[width=2.2in]{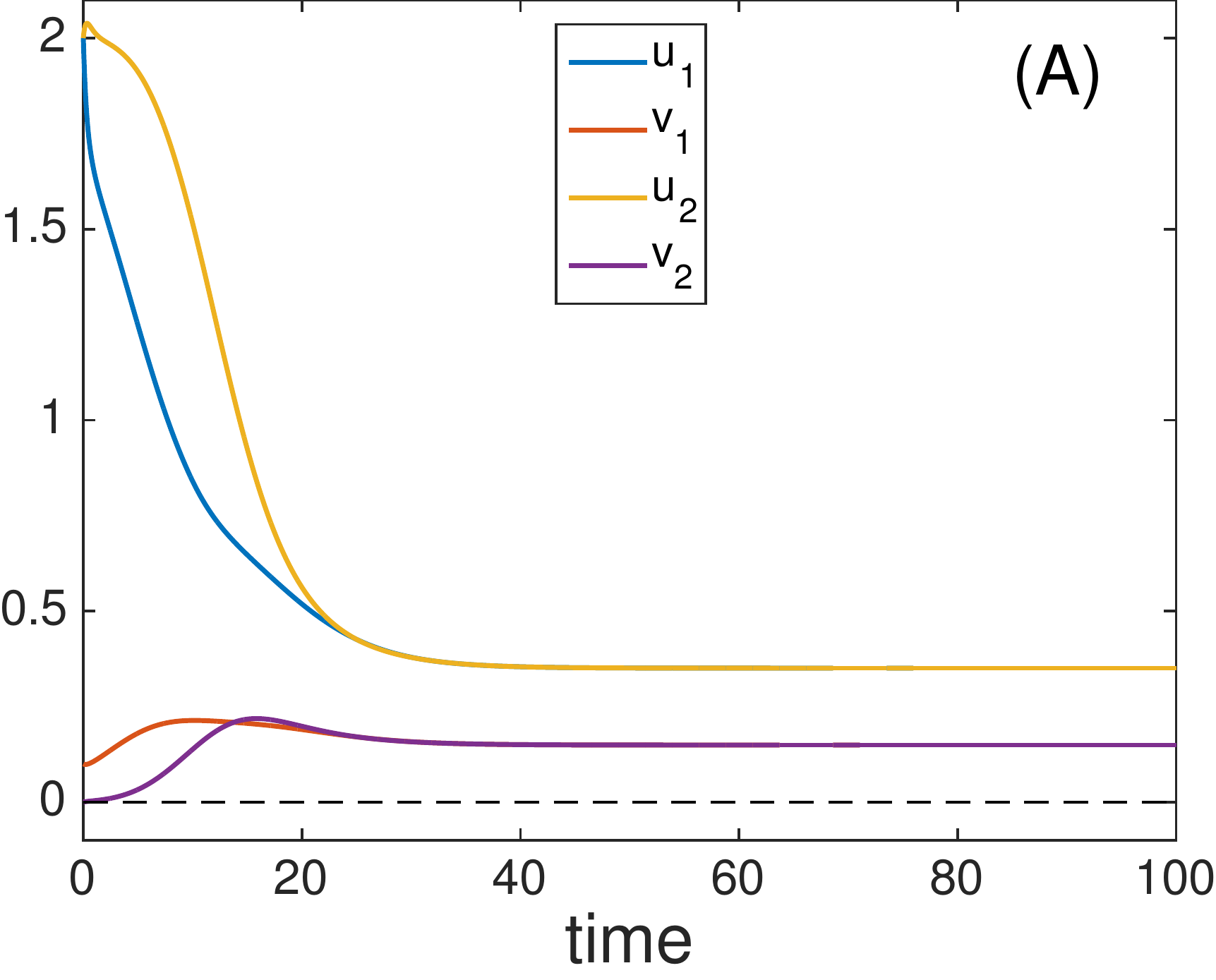} \includegraphics[width=2.2in]{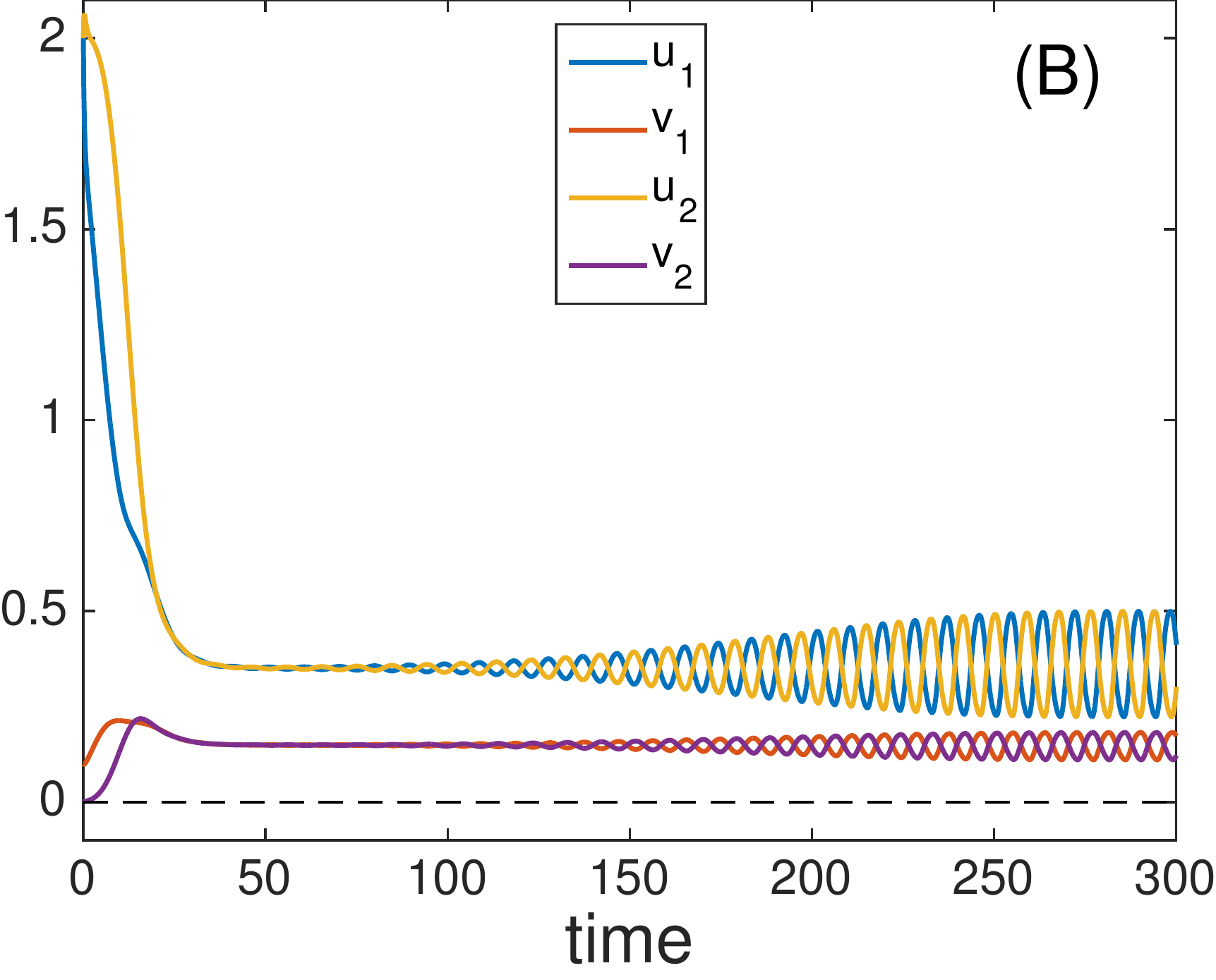}\\
\includegraphics[width=2.2in]{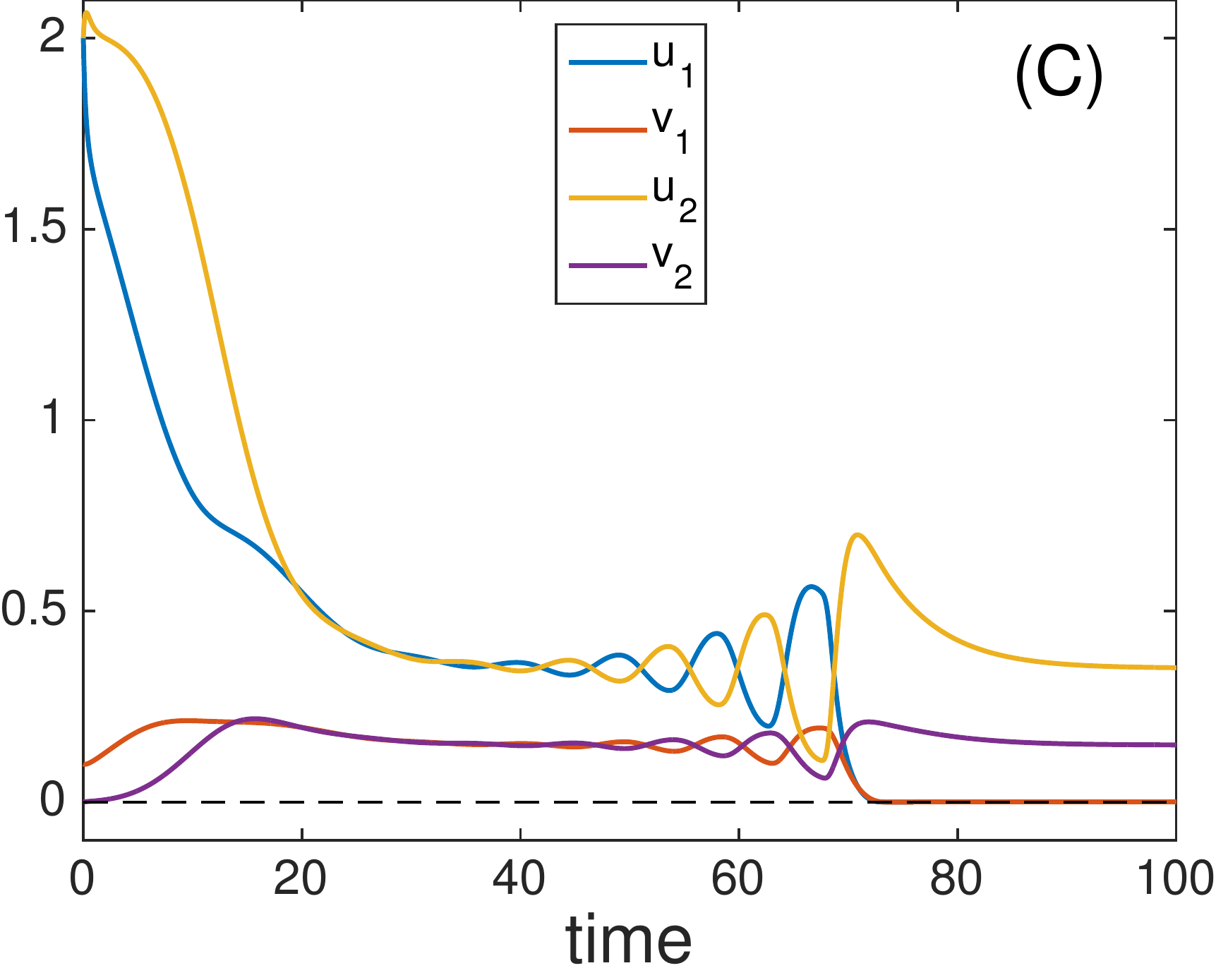} \includegraphics[width=2.2in]{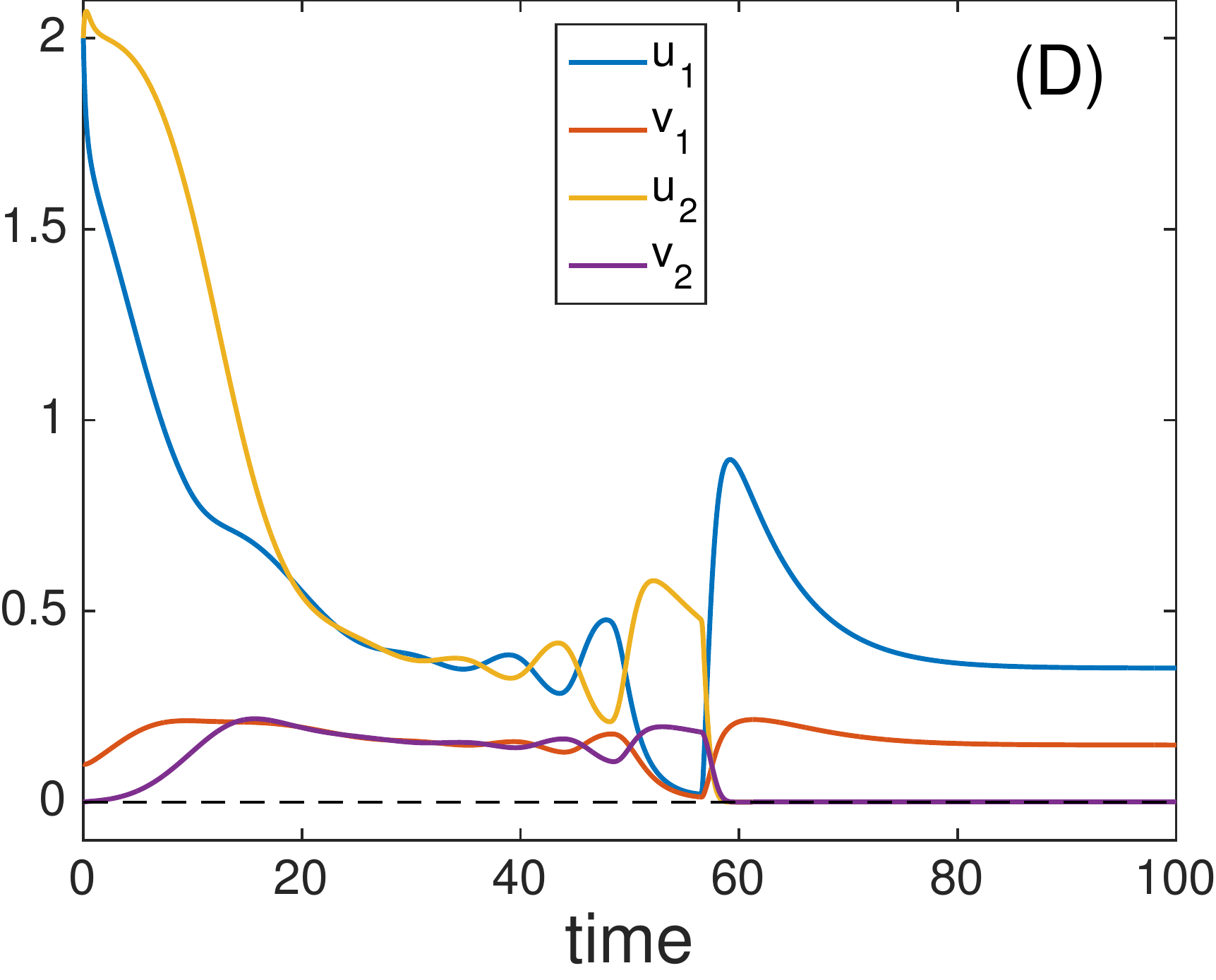}

}
\caption{Time series over varied $\beta_u$ with $\beta_v=0.2$ and initial conditions $u_1(0)=2, u_2(0)=2, v_1(0)=0.1, v_2(0)=0$ fixed. A. $\beta_u=0.5$, corresponding to the circle in Figure \ref{fig:betaBif}. The two nests both reach their interior equilibrium point. B. $\beta_u=0.95$, corresponding to the square in Figure \ref{fig:betaBif}. The interior equilibrium has undergone a Hopf bifurcation and the two worker and cheater populations oscillate in anti-phase between the two nests. C. $\beta_u=1.1$, corresponding to the upward facing triangle in Figure \ref{fig:betaBif}. The interior equilibrium has lost stability completely through a torus bifurcation, and nest 1 collapses. D. $\beta_u=1.2$, corresponding to the downward facing triangle in Figure \ref{fig:betaBif}. The interior equilibrium is unstable, but now nest 2 collapses. Increasing $\beta_u$ further with $\beta_v$ fixed will change which nest collapses.}
\label{fig:trajectories}
\end{figure}

Figure \ref{fig:trajectories} shows the generic behaviors of System (\ref{eq:twoNest1}) for $(\beta_u,\beta_v)$ pairs fixed in each region in Figure \ref{fig:betaBif}. The cheater motility is fixed at $\beta_v=0.2$ in each subfigure, and the system is initialized with both worker populations at their cheater-free steady state values, $u_1(0)=u_2(0)=2$, and a small perturbation of cheaters are introduced into nest 1 only: $v_1(0)=0.1, v_2(0)=0.$ In Figure \ref{fig:trajectories}A, worker motility is chosen as $\beta_u=0.5$, corresponding to the black circle  in Figure \ref{fig:betaBif}, and the solution tends to the coexisting steady state $T^*$. Figure \ref{fig:trajectories}B shows the solution curves with $\beta_u=0.95$, corresponding to the square in the grey region of Figure \ref{fig:betaBif}. Both nests persist for all time, but their respective worker and cheater populations oscillate in antiphase. Figures \ref{fig:trajectories}C and D both show solution examples from the blue region in Figure \ref{fig:betaBif}. In \ref{fig:trajectories}C, $\beta_u=1.1$, and nest 1 collapses while nest 2 tends to its coexisting steady state. In \ref{fig:trajectories}D, $\beta_u$ is increased to $1.2$, and now nest 2 collapses while both the workers and cheaters persist in nest 1. This suggests that the system is very sensitive to changes in motility rates, 
%within the grey region in Figure \ref{fig:betaBif}. We 
which we investigate below.

%%%%%%%%%%%%%%%%%%%%%%%%%%%%%%%%%%%%%%%%%
%%%%%%%%%%%%%%%%%%%%%%%%%%%%%%%%%%%%%%%%%

The implications of Figures \ref{fig:betaBif} and \ref{fig:trajectories} are somewhat surprising: if workers migrate too quickly to the nest with higher fitness, they cause the other nest to collapse. Increased motility of these contributors therefore comes at a fairly high cost. Increased motility of the cheaters, on the other hand, is strictly beneficial in the sense of preservation of the network, even though the presence of cheaters decreases the total population within each colony.

\begin{figure}[!t] % H
{\centering
\includegraphics[width=3.8in]{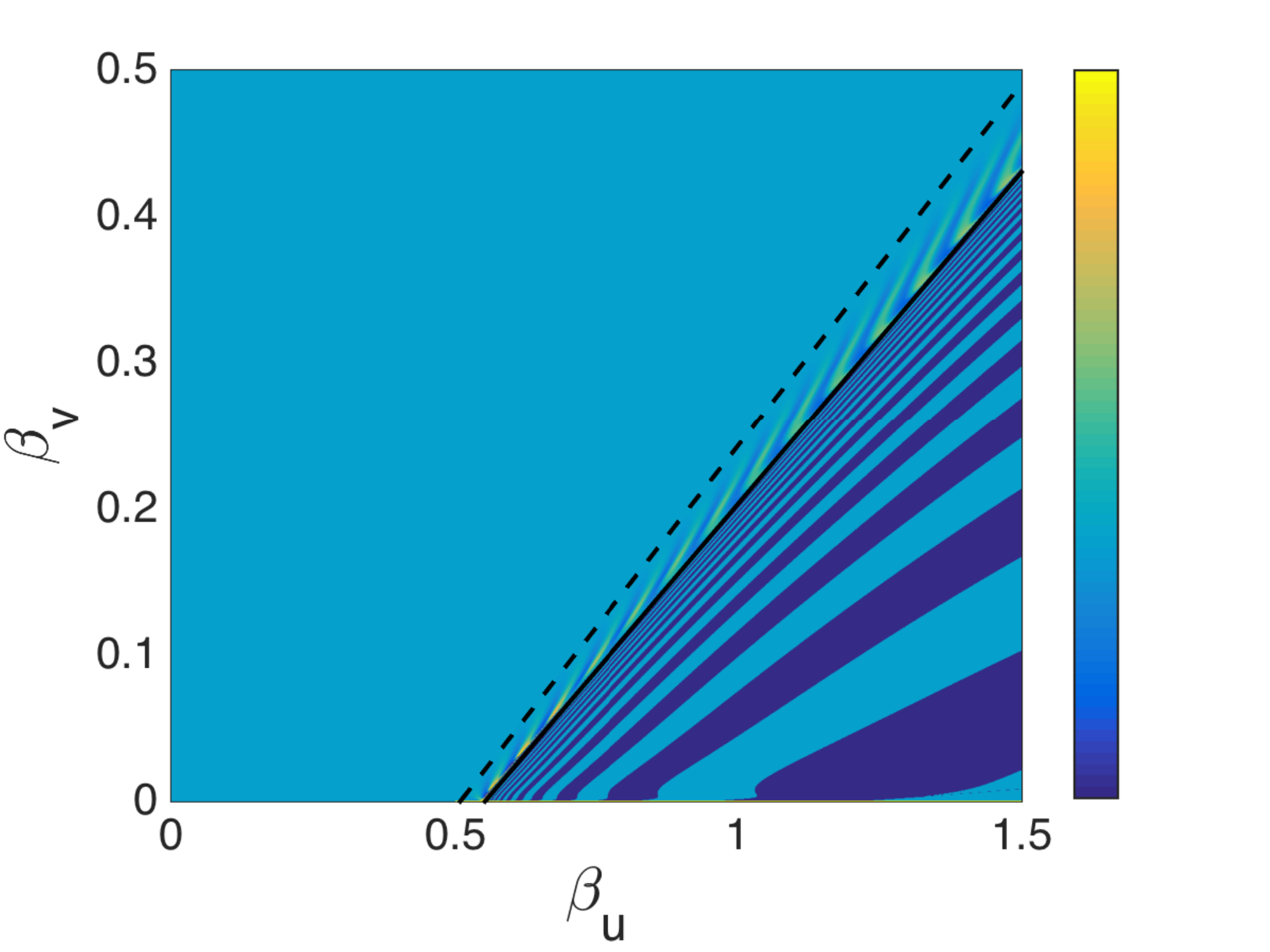}

}
\caption{Basins of attraction across varied motility rates for two nests of equal quality. Initial conditions are fixed at $u_1(0)=2, v_1(0)=0.1, u_2(0)=2, v_2(0)=0$. {The color corresponds to the size of the worker population in nest 1 at time $t=150$; that is, the lighter the color, the larger the population in nest 1. The dashed and solid lines correspond to the lines of Hopf and torus bifurcations, respectively, as shown in Figure \ref{fig:betaBif}.} To the left of the Hopf bifurcation, both nests maintain a stable population of coexisting workers and cheaters. Between the line of Hopf bifurcations and that of torus bifurcations, the individuals oscillate between the two nests. To the right of the torus bifurcation, one of the two nests necessary collapses: dark blue corresponds to nest 1 collapsing; light blue corresponds to nest 2 collapsing.}
\label{fig:betaBasin}
\end{figure}

Figures \ref{fig:trajectories}C and D show that, in the parameter regime beyond the torus bifurcation, the nest that ultimately survives is highly dependent on the motility rates of the two populations. This is verified by Figure
\ref{fig:betaBasin}: after the torus bifurcation, ``bands'' of survival appear, in which the two nests alternately survive and collapse as the motilities vary (compare  Figure \ref{fig:basinSym}). 
%Further, for any fixed $(\beta_u,\beta_v)$ pair, the surviving nest varies over initial population composition (Figure \ref{fig:diffBasins}). This implies that the nest that ultimately collapses is not predictable in general, but, importantly, one of the two nests will necessarily collapse if $\beta_u$ is sufficiently large.

\subsection{Colonies of Unequal Quality}
\begin{figure}[!t] %H 
{\centering
\includegraphics[width=2in]{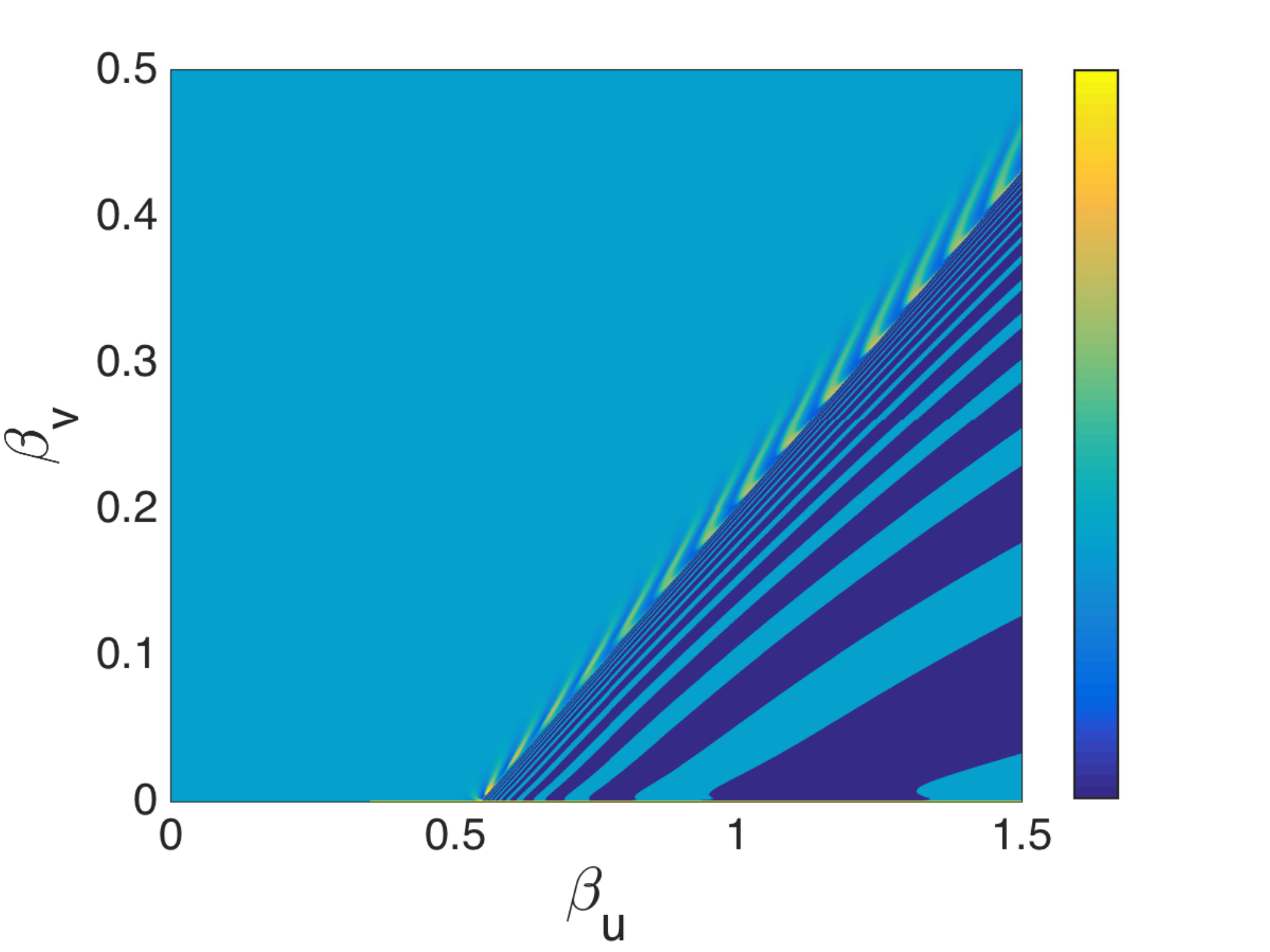} \includegraphics[width=2in]{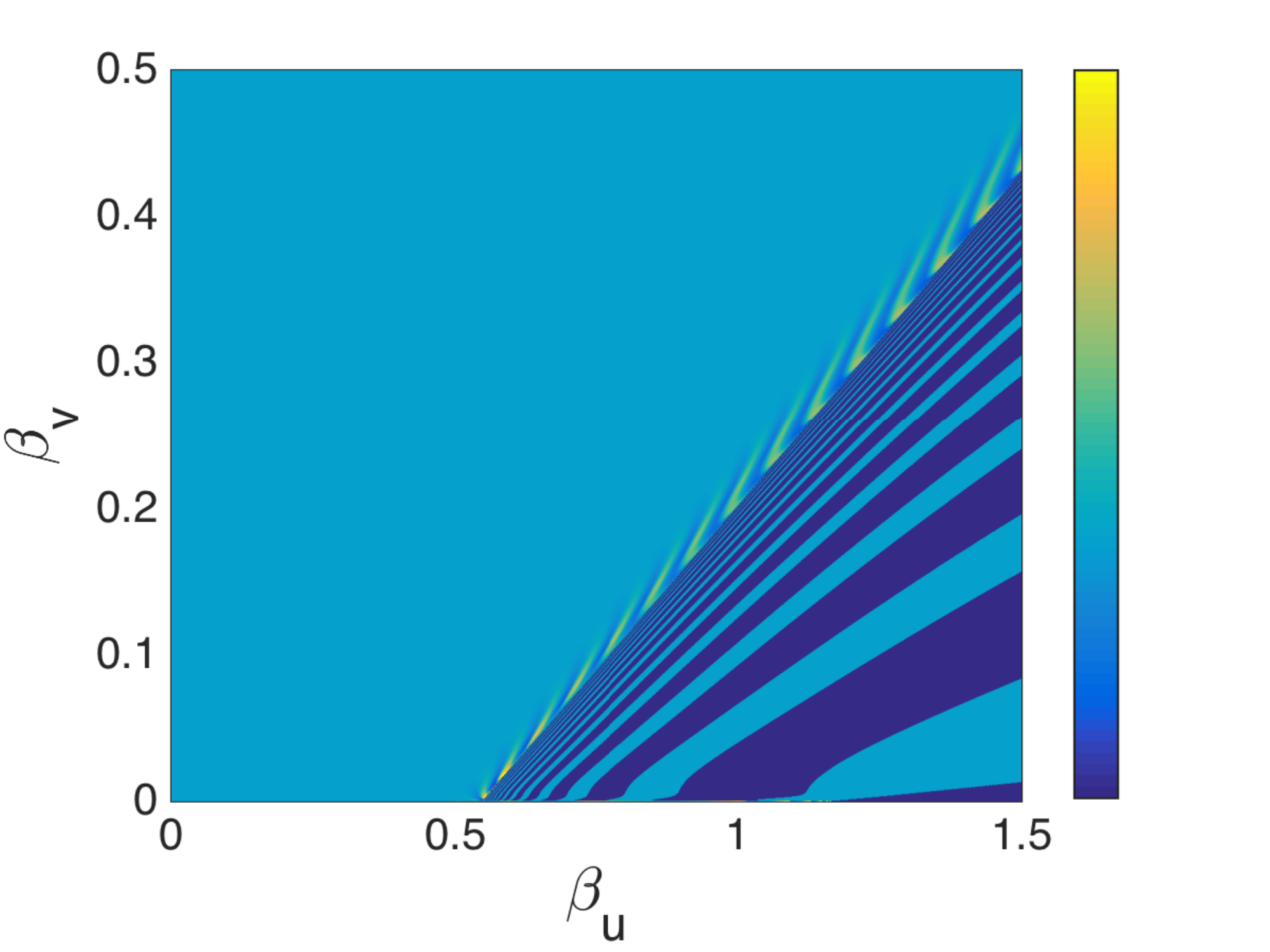}

\includegraphics[width=2in]{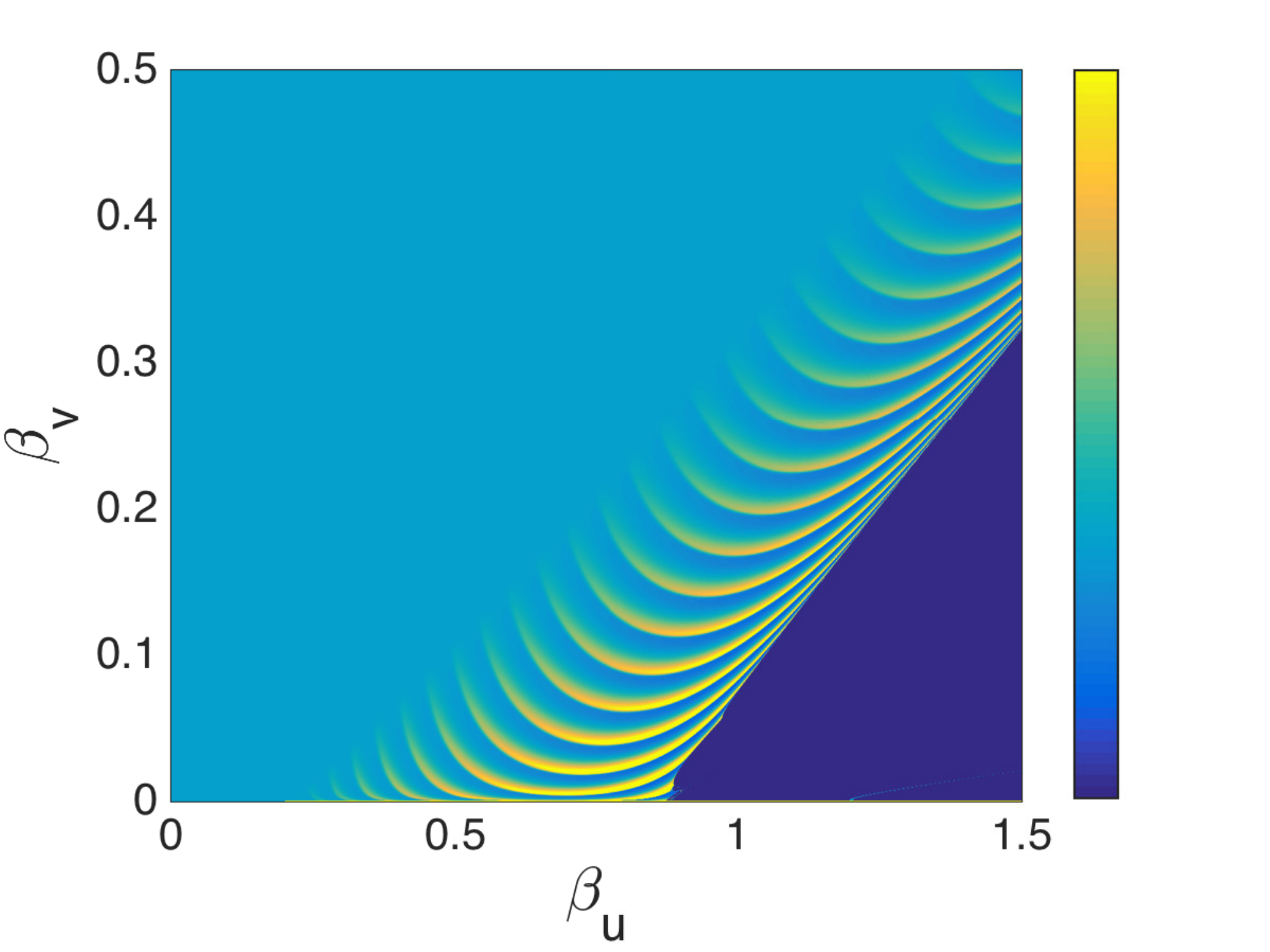} \includegraphics[width=2in]{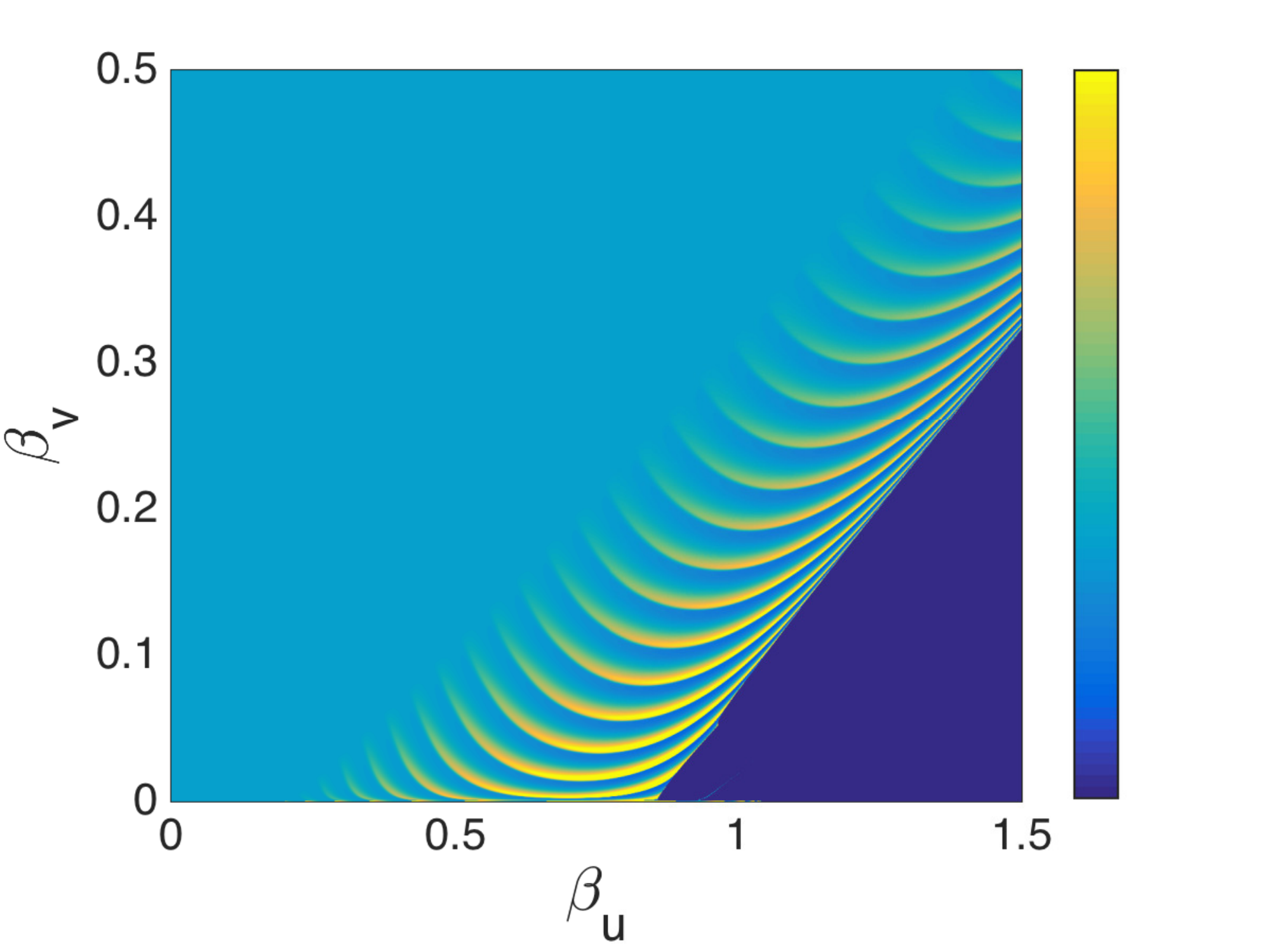}

}

\caption{Basins of attraction over varied motility rates for two nests of unequal quality. Initial conditions are fixed at $u_1(0)=2, v_1(0)=0.1, u_2(0)=2, v_2(0)=0$. The colors are the same as those in Figure \ref{fig:betaBasin}. A. The quality of nest 1 is 50\% greater than that of nest 2: $\gamma_1=0.75$ and $\gamma_2=1$. The qualitative behavior of the system over varied motility rates remains largely unchanged compared to the equal nest quality case. B. The quality of nest 1 is ten times greater than that of nest 2: $\gamma_1=0.1$ and $\gamma_2=1$. In the region to the right of the torus bifurcation, the lower quality nest almost always is the one to collapse.}
\label{fig:betaBasinsUnequal}
\end{figure}

The above analysis only holds when considering colonies comprised of two nests of equal quality ($\gamma_1=\gamma_2$). When the quality of the nests is allowed to differ, the question of stability is complicated by the loss of symmetry over the switching manifolds. While there is an emerging theory on classifying the stability of piecewise smooth systems (see, e.g., \cite{Harris,Leine}), the nonlinearity of the switching manifolds in System (\ref{eq:twoNest1}) renders this analysis unfeasible. We instead explore the behavior of such a system numerically. Figure \ref{fig:betaBasinsUnequal}A shows the basin of attraction across varied $\beta_u, \beta_v$ pairs when $\gamma_1=0.75$ and $\gamma_2=1$.  The qualitative behavior remains unchanged: for $\beta_u$ sufficiently small, the two nests persist with both worker and cheater populations at their respective steady state values. As $\beta_u$ is increased, the system undergoes a Hopf bifurcation, after which the populations periodically migrate between the two nests. Continuing to increase $\beta_u$ causes the periodic limit cycle to lose its stability, and one of the nests necessarily collapses. In this parameter region, the basins of attraction of the two nests form ``bands,'' similar to the equal quality case (Figure \ref{fig:betaBasin}). We conclude that if the nests are similar enough in quality, their dynamics reflects that of the equal-quality case.

If we allow the quality of the two nests to vary greatly, however, the lower quality nest will typically collapse once the two-nest system becomes unstable. Figure \ref{fig:betaBasinsUnequal}B shows the basins of attraction in the $\beta_u,\beta_v$ plane when $\gamma_1=0.1$ and $\gamma_2=1.$ The system once again passes through through a Hopf bifurcation as $\beta_u$ is increased through a critical threshold, and the resulting limit cycle still loses stability through a torus bifurcation as $\beta_u$ increases further. However, in the unstable parameter region, the higher quality nest (nest 1, in this case) almost always survives, while the lower quality nest (nest 2) almost always collapses.

Interestingly, there is a small band of $\beta_u,\beta_v$ pairs such that the lower quality nest survives while the higher quality nest collapses, though that band disappears as the discrepancy in quality grows even greater. We note here that this behavior seems to generalize for colonies of larger size: if the discrepancy in quality between each colony in a network is small, the system behaves similarly to equal-quality case. If one or more nests are of significantly higher quality, then those nests are considerably less likely to collapse once the coexisting state of the colony network losing stability. Because of this seemingly general behavior, we will only consider networks comprised of colonies of equal quality, and not present any further results on the effects of unequal quality.

\section{N colony system}\label{sec:nNest}

We now consider a network of $N$ colonies, with migration possible between any two of them; that is, the network of nests is all-to-all connected. To do so, we return to the $2N$-equation system from Section \ref{sec:model}
\begin{equation}\tag{\ref{eq:nNest}}
\begin{aligned}
\dot u_i &=  \displaystyle\sum_{j=1}^N \alpha_{ij}H_u(u_i,v_i,u_j,v_j)+u_iF_i(u_i,v_i)\\
\dot v_i &=  \displaystyle\sum_{j=1}^N \beta_{ij}H_v(u_j,v_j,u_2,v_j)+v_iG_i(u_i,v_i),
\end{aligned}
\end{equation}
 for $i=1\dots N$. We will only consider the case in which $\alpha_{ij}=\beta_u$ and $\beta_{ij}=\beta_v$ for each $1\leq i,j\leq N$; that is, all workers migrate at the same rate, as do all cheaters.
Just as in the two-nest case, we aim to study conditions that generically lead to persistence or collapse. We only consider the symmetric case in which $\gamma_i=\gamma_j=\gamma$ for each $1\leq i,j\leq N$, and consequently $F_i=F$ and $G_i=G$ for each $i$. Under this assumption, we can again exploit the symmetry of the system over each switching manifold, and apply linear stability analysis to determine the structure of the coexisting state.

The Jacobian matrix of the $N$-nest coexisting equilibrium point is the $2N\times 2N$ matrix
\[
J_N=\begin{bmatrix}
    P+(N-1)Q & -Q & -Q & \dots  & -Q \\
    -Q & P+(N-1)Q & -Q & \dots  & -Q \\
    -Q & -Q & P+(N-1)Q & \dots  & -Q\\
    \vdots & \vdots & \vdots & \ddots  & \vdots \\
    -Q & -Q & -Q & \dots  & P+(N-1)Q
\end{bmatrix},
\] where 

\[
P=\begin{bmatrix}
    \frac{\partial F}{\partial u} u^* & \frac{\partial F}{\partial v} u^* \\
    \frac{\partial G}{\partial u} v^* & \frac{\partial G}{\partial v} v^*
\end{bmatrix},
\]
and
\[
Q=\begin{bmatrix}
    \beta_u\frac{\partial F}{\partial u} u^* & \beta_u\frac{\partial F}{\partial v} u^* \\
    \beta_v\frac{\partial G}{\partial u} v^* & \beta_v\frac{\partial G}{\partial v} v^*
\end{bmatrix}.
\]
The characteristic polynomial of $J_N$ can be shown to be
\begin{equation}\label{eq:charPoly}
\chi_{J_N}(x)=\chi_{F}(x)\left[\chi_{P+NQ}(x)\right]^{N-1},
\end{equation} 
where $\chi_{P}(x)$ and $\chi_{P+NQ}(x)$ are the characteristic polynomials of $P$ and $P+NQ$, respectively. The stability of the coexisting state is therefore determined by  the eigenvalues of these two $2\times2$ matrices. The matrix $P$ is the Jacobian matrix of System (\ref{eq:noMigEps}) evaluated at the interior equilibrium point E$_3$, and has eigenvalues with negative real part as long as condition (\ref{eq:E3stab}) is satisfied.

\begin{figure}[H]
{\centering
\includegraphics[width=3in]{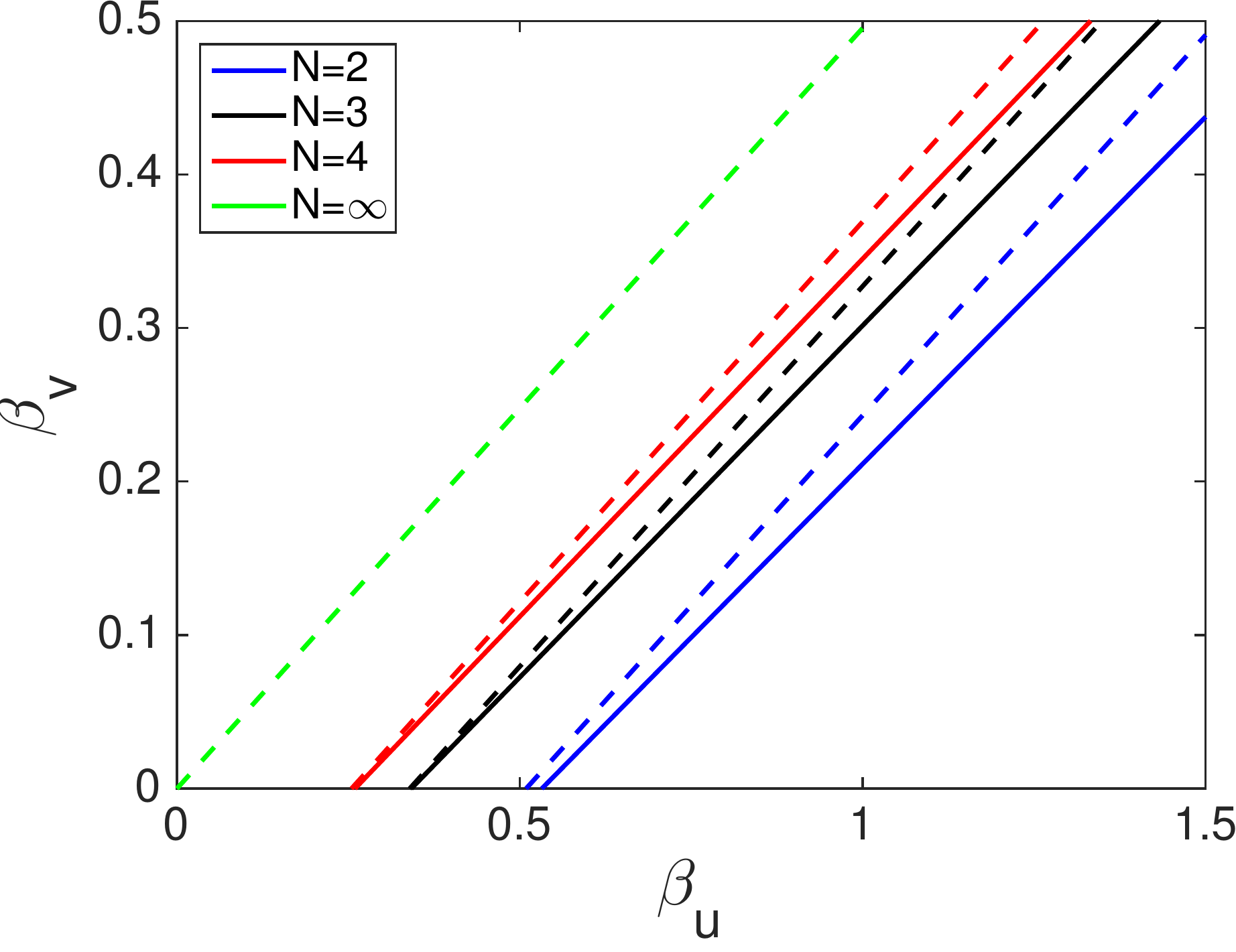}

}
\caption{Hopf and torus bifurcations in colony networks of various sizes. For each $N$, the dashed line corresponds to a line of Hopf bifurcations, defined by $\sigma_N=0$, and the solid line corresponds to a line of torus bifurcations, found numerically. As worker motility is increased, the stability of each $N$-nest system is lost, causing a single nest to collapse as the system undergoes each torus bifurcation, until only a single nest remains.}
\label{fig:multiBif}
\end{figure}

As long as the one-nest system is stable, then the stability of the $N$ nest system is determined by the matrix $P+NQ$. This matrix generally has a complex pair of eigenvalues with real part
\begin{equation*}
\sigma_N=\frac{1}{2}(1+N\beta_u)u^*\left(\frac{r_ucv^*}{(u^*+v^*)^2}-\gamma\right)-\frac{1}{2}(1+N\beta_v)v^*\left(\frac{r_vcu^*}{(u^*+v^*)^2}+\gamma\right).
\end{equation*}
The $N=2$ case was presented in Section \ref{sec:twoNest} (Figure \ref{fig:betaBif}). For general $N$, when $\beta_u=\beta_v=0$, $Q$ is the zero matrix, and the eigenvalues of $P+NQ=P$ consequently have negative real part. By continuity, $\sigma_N$ must be negative for $\beta_u$ and $\beta_v$ small. Increasing $\beta_u$, however, will generically cause $\sigma_N$ to become positive, upon which the equilibrium goes through a Hopf bifurcation. The level sets $\sigma_N=0$ for $N=2,3,$ and $4$ are plotted as the dashed lines in Figure \ref{fig:multiBif}. The solid lines following each line of Hopf bifurcations corresponds to a line of torus bifurcations, found numerically, through which the colony network of size $N$ loses its stability. The behavior mimics that of the two-nest system: if worker motility is small relative to cheater motility, the network is stable, but as workers become faster, it begins to lose stability. As worker motility is increased, the network collapses colony by colony until only a single one  remains.  Each level set defines a line in the $\beta_u$, $\beta_v$ plane, the slope $m$ of which is independent of $N$. As $N\to\infty$, the line defined by $\sigma_N=0$ converges to the line passing through the origin with slope $m$, given by the green dashed line in Figure \ref{fig:multiBif}. Above and to the left of this line, a colony network of any size can be supported.

%\begin{itemize}
%\item Stability is determined by the eigenvalues of $F$ and $F+NH$\checkmark
%\item $F$ is the Jacobian of system (\ref{eq:noMig}) $\rightarrow$ stability determined by conditions discussed in section 2\checkmark
%\item $F+NH$ in general has complex eigenvalues, with real part equal to one half the trace of $F+NH$ $\rightarrow$ compare to 2 nest case
%\item make figs for the three nest case similar to those in figures 2 and 3
%\item explore nests that aren't all to all connected--``The above results are for all to all connected colonies; similar results hold for different network structures, though the bifurcation curves are more difficult to compute analytically. For example, \{a ring, maybe small word or something, etc\}
%\end{itemize}

\begin{figure}[H]
{\centering

\includegraphics[width=1.19in]{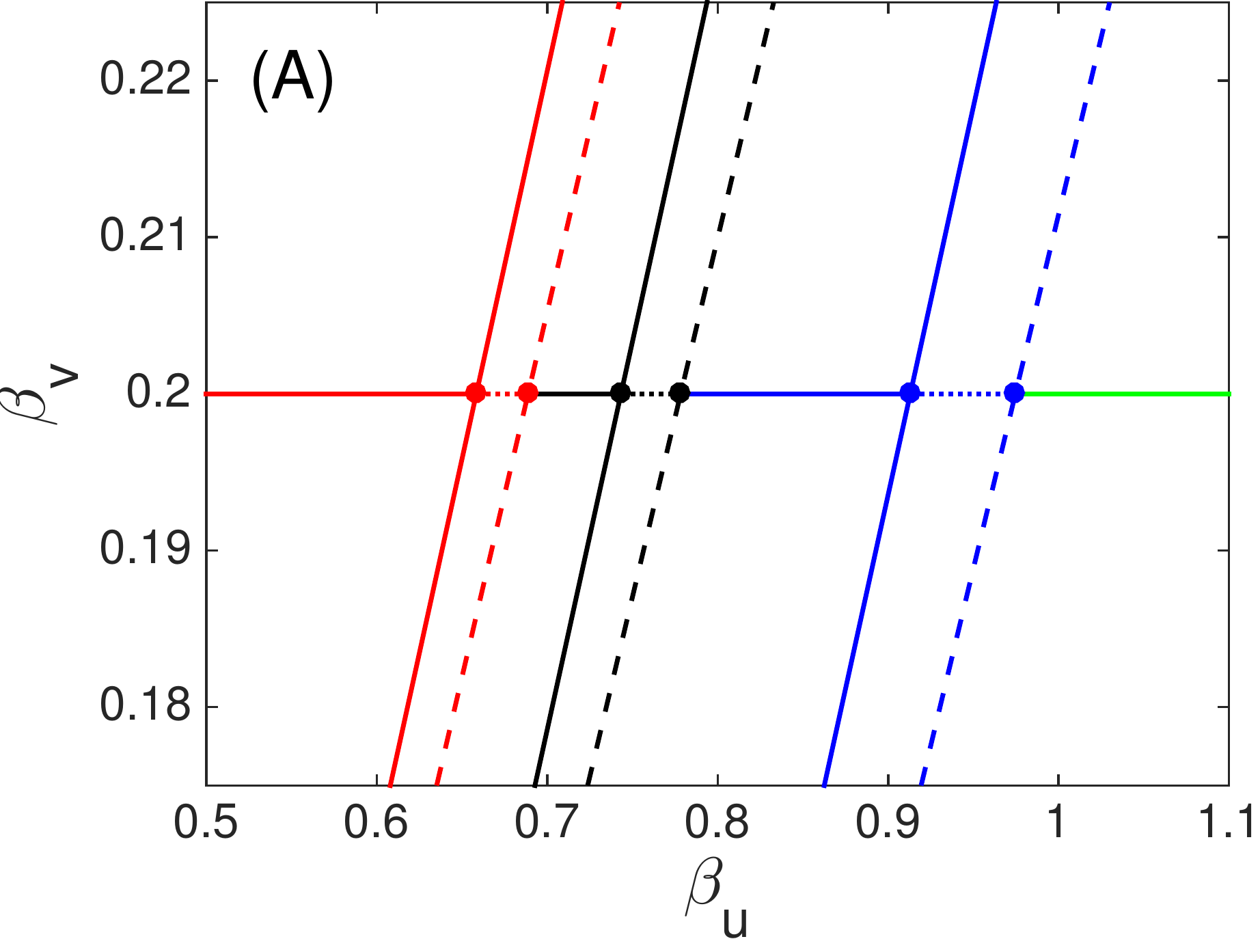} \includegraphics[width=1.1in]{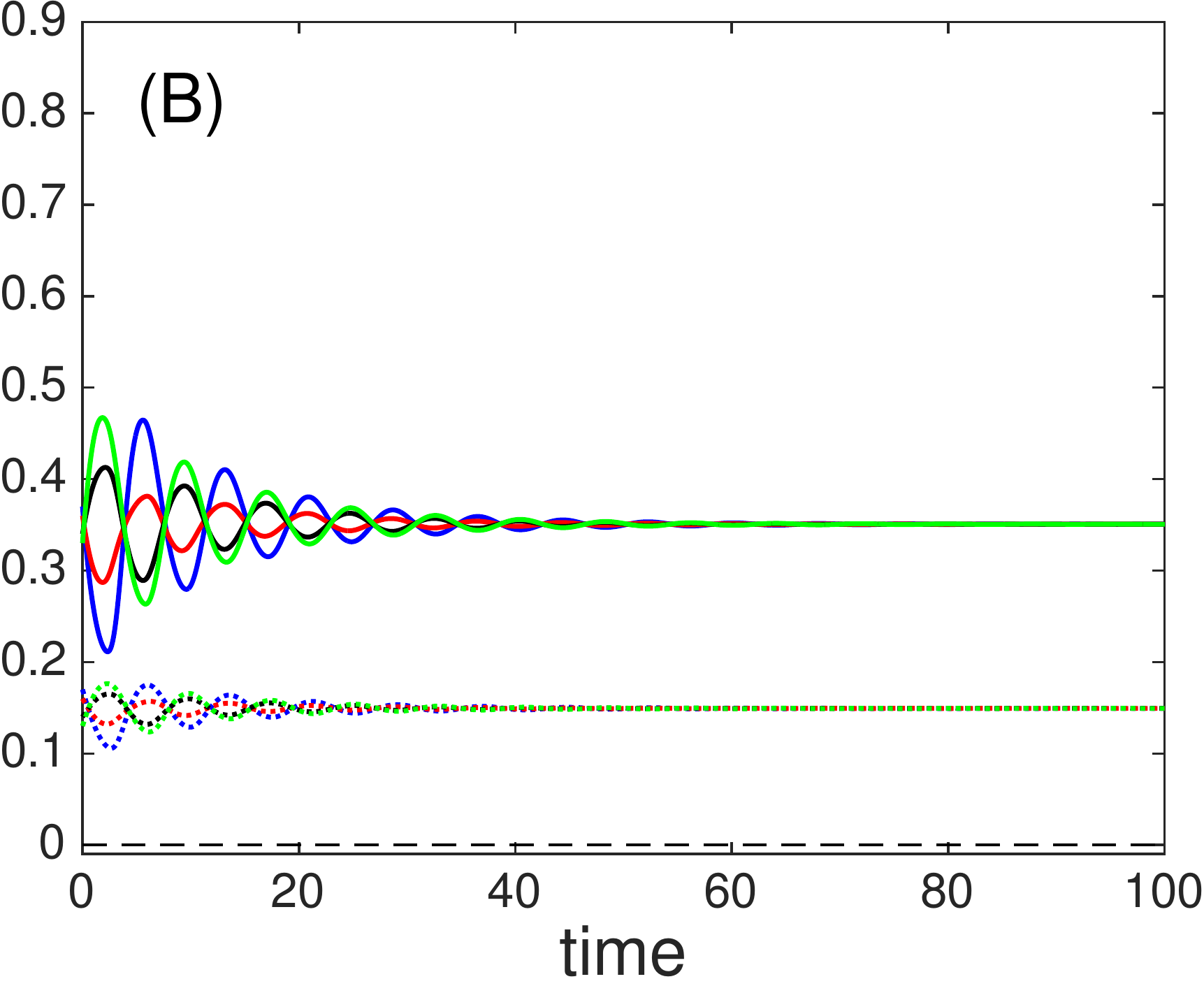} \includegraphics[width=1.1in]{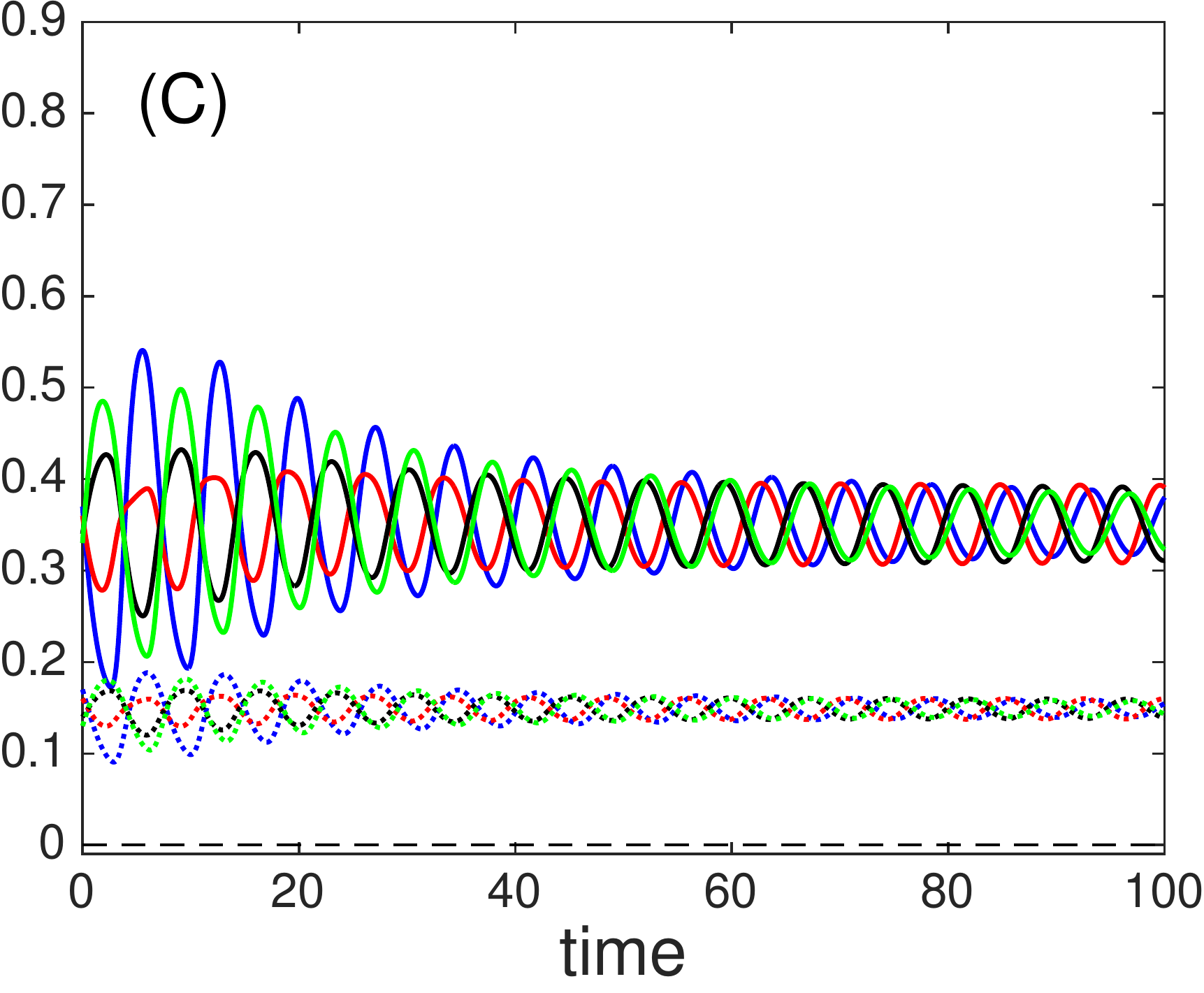} \includegraphics[width=1.1in]{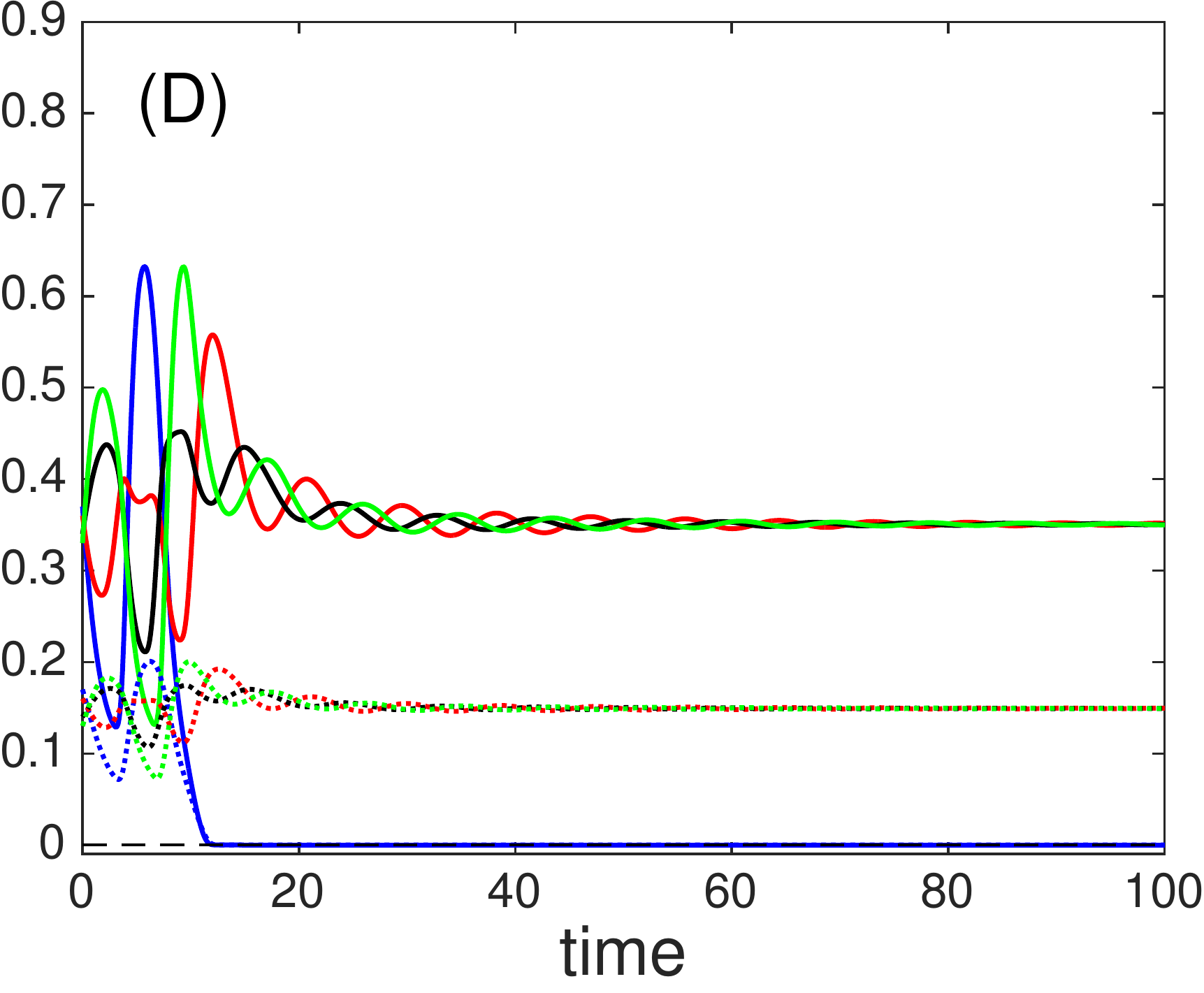}\\
\includegraphics[width=1.1in]{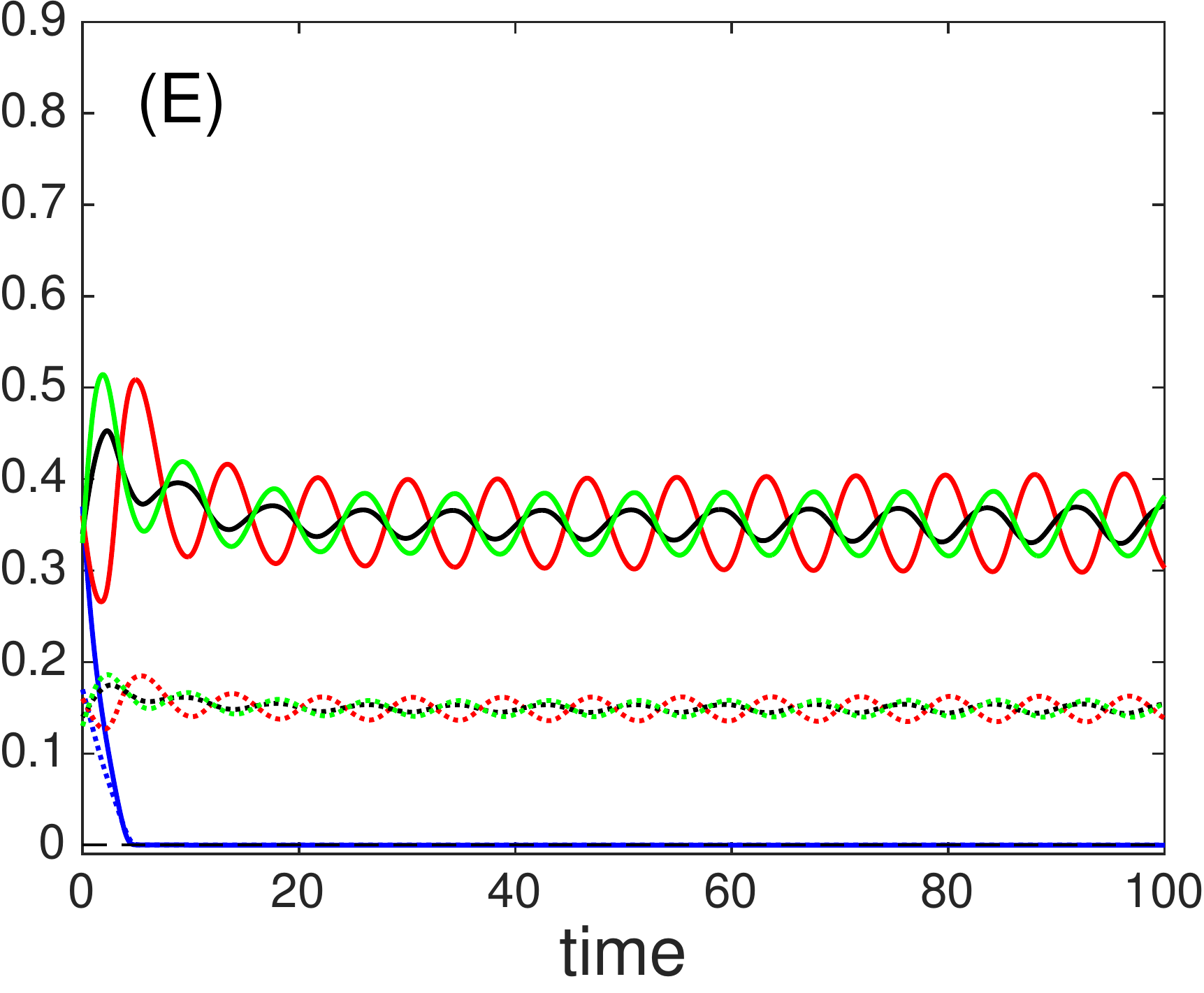} \includegraphics[width=1.1in]{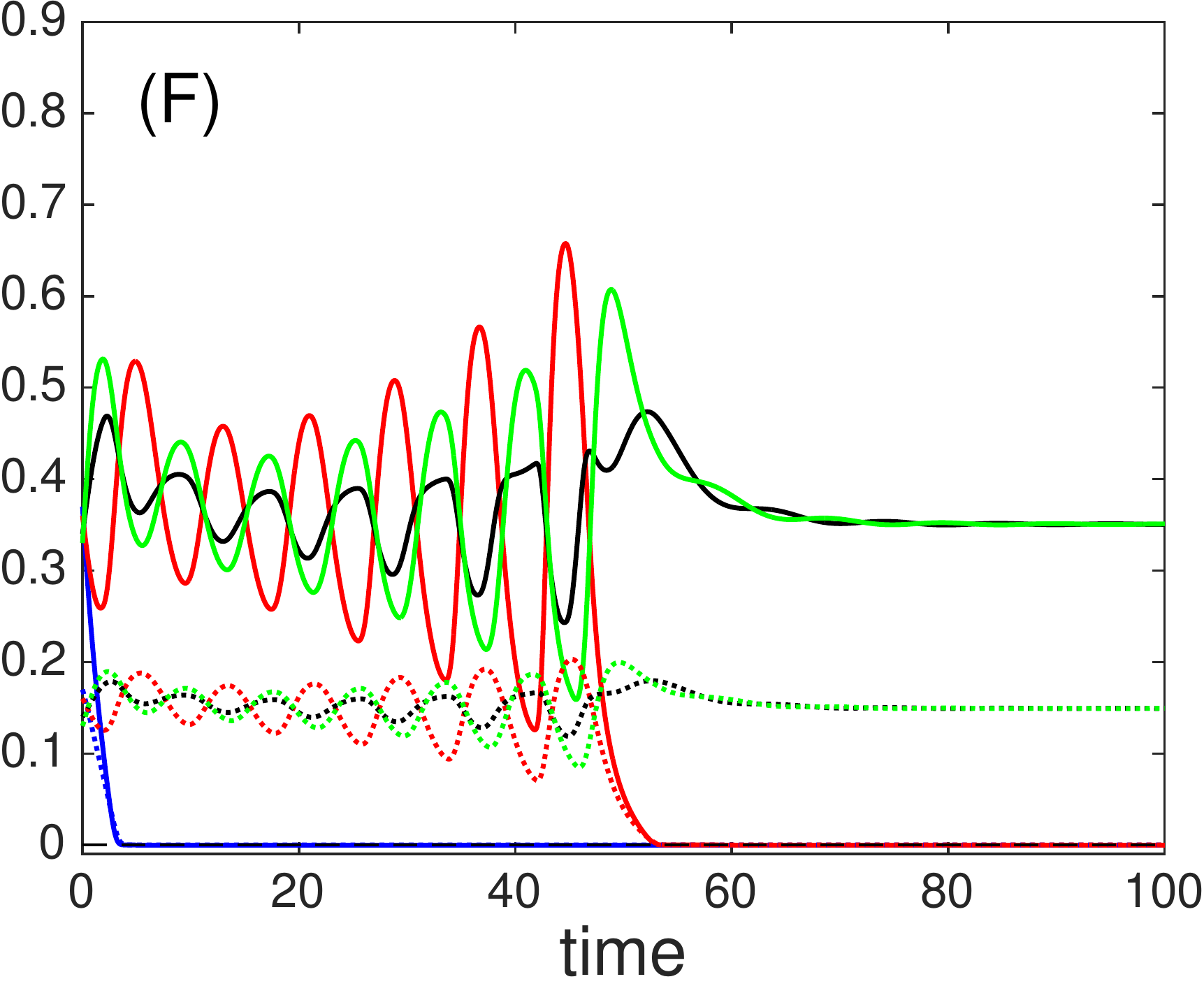} \includegraphics[width=1.1in]{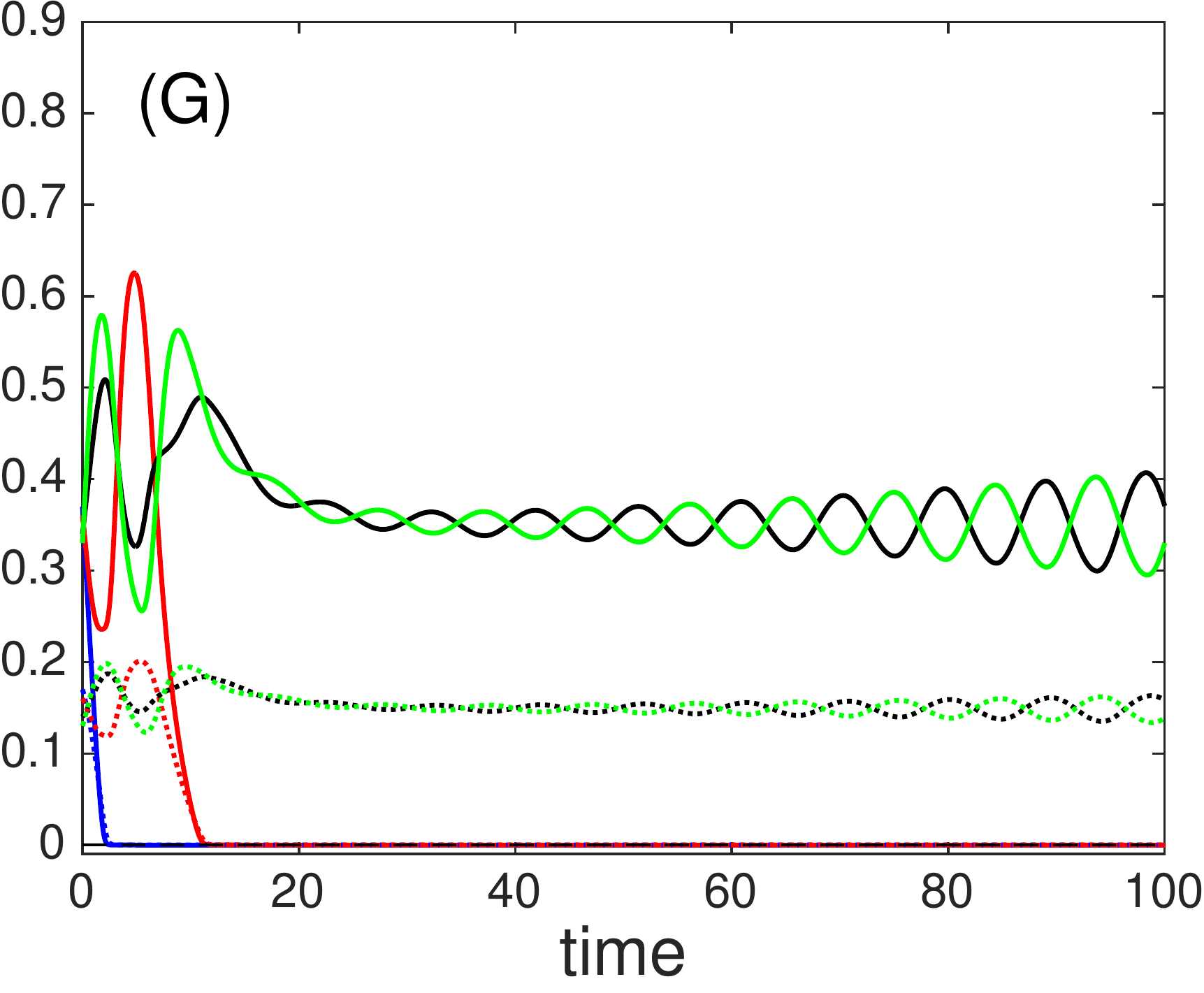} \includegraphics[width=1.1in]{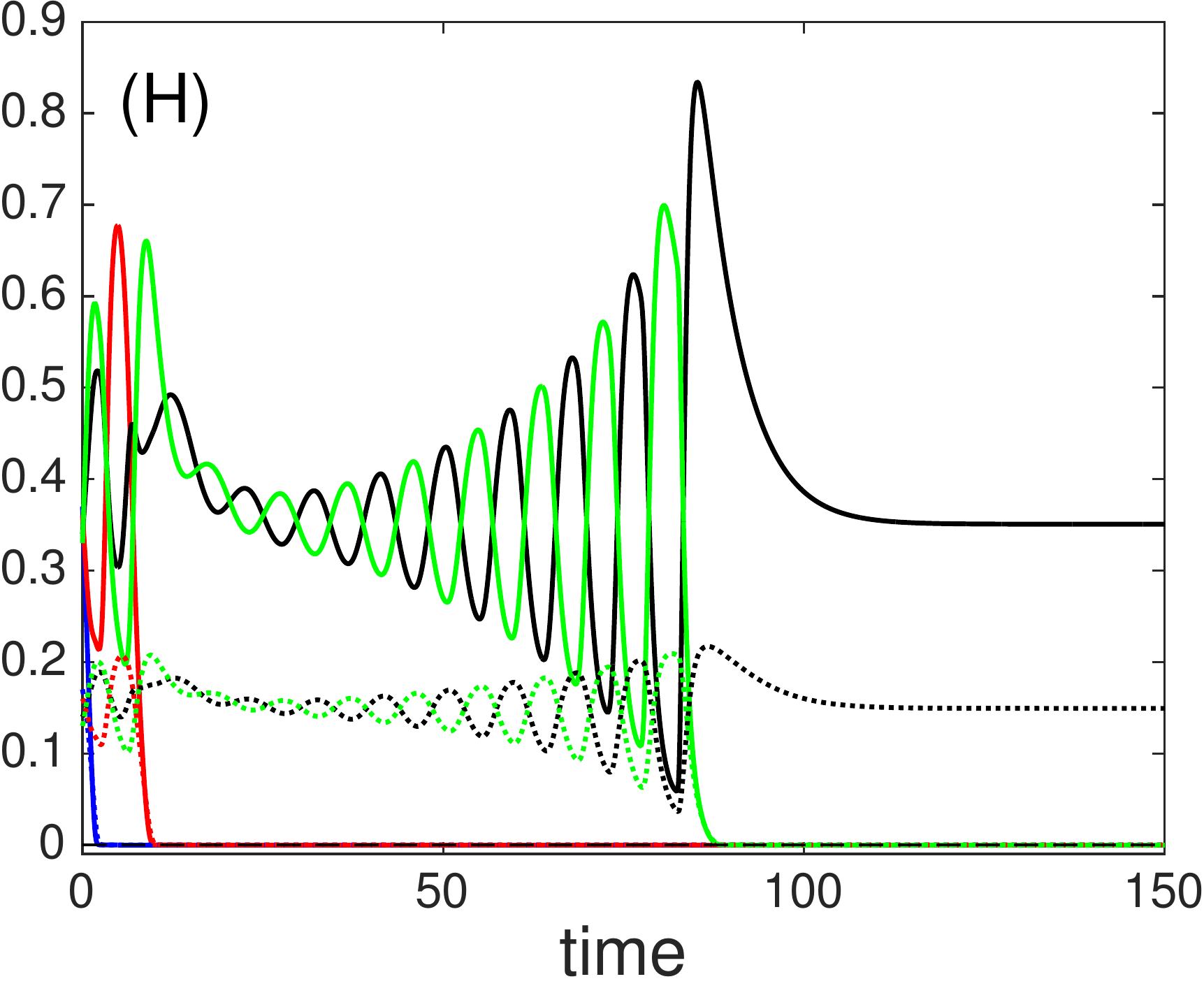}

}
\caption{Example trajectories after each bifurcation in a four-nest system. The system was initialized by assigning $u_i=0.35+u_\epsilon$ and $v_i=0.15+v_\epsilon$, where $u_\epsilon$ was selected at random from the interval $[-0.1,0.1]$ and $v_\epsilon$ from $[-0.06,0.06]$. For each simulation, $\beta_v=0.2$ was fixed, while $\beta_u$ was increased to a value past the next successive bifurcation. In panels (B)-(H), the $\beta_u$ values used were, respectively, 0.6, 0.66, 0.7, 0.75, 0.8, 0.96, and 1.0.}
\label{fig:severalBifs}
\end{figure}

For fixed $\beta_v$, the stability of the $N, N-1,\dots,2$ nest systems are lost in sequence through torus bifurcations as $\beta_u$ is increased. This suggests that for fixed migration rates $\beta_u$ and $\beta_v$, there is a maximum number of colonies that a network can support. Figure \ref{fig:severalBifs} shows example trajectories of a four colony, all-to-all connected system, in which each colony is of the same quality; $\beta_u$ is varied while $\beta_v=0.2$ fixed. For small $\beta_u$, the system supports stable equilibria within each of the four nests (Figure \ref{fig:severalBifs}B). As $\beta_u$ is increased, the four nest system undergoes a Hopf bifurcation, and the individuals actively oscillate between each of the four nests (Figure \ref{fig:severalBifs}C). Increasing $\beta_u$ further causes the stability of the four nest system to be completely lost, and one of the nests collapses (Figure \ref{fig:severalBifs}D). From here, the network is effectively a three colony system, and increasing $\beta_u$ causes the three remaining nests to similarly undergo a Hopf bifurcation, after which the population within each nest oscillates (Figure \ref{fig:severalBifs}E) and then a torus bifurcation, after which another nest collapses, leaving only two active nests (Figure \ref{fig:severalBifs}F). The system is then identical to the two-nest system studied in Section \ref{sec:twoNest}, which once again undergoes a Hopf bifurcation (Figure \ref{fig:severalBifs}G), then a torus bifurcation as $\beta_u$ is increased, resulting in only a single nest surviving (Figure \ref{fig:severalBifs}H). It is important to note that, much like in the two-nest system, the nests that collapse are effectively random. This is in part due to the fact that each nest in Figure \ref{fig:severalBifs} has the same carrying capacity and is connected to each of the other three nests, though the simulations shown in Figure \ref{fig:betaBasinsUnequal} suggest that even when the nests are of unequal quality, predicting which nests will collapse is likely impossible. Moreover, the nests that ultimately collapse for fixed $\beta_u$ and $\beta_v$ are determined entirely by initial conditions, and small changes in initial conditions can result in entirely different subsets of an $N$-colony network collapsing, further complicating any predictions (see Figure \ref{fig:basinSym}).

 \subsection{Dependence on network structure}
 
The above results hold for all-to-all connected networks of nests. Of course, nests in a natural setting are not necessarily arranged in such a way that movement between any given pair is possible. We therefore now consider colonies connected via more complicated network structures. While generalizing the above results to a network with arbitrary structure is infeasible, we instead focus on two simple networks that lend themselves to tractable analysis to provide evidence that the stabilizing ability of fast cheaters appears to be network-independent. We also briefly consider a network in which connections between colonies are randomly generated. In each case considered, the qualitative behavior of the system matches that of an all-to-all connected network: the coexisting state is stable for small $\beta_u$, but nests begin to collapse and the system begins to break down as this motility is increased.  Unlike the all-to-all connected cases, however, a more general network structure can fragment into two or more independent networks.
 
 \subsubsection{Two connected networks}\label{sec:twoColonies}
 
\begin{figure}[H]
{\centering
\hspace{0.6cm}
\includegraphics[width=1.7in]{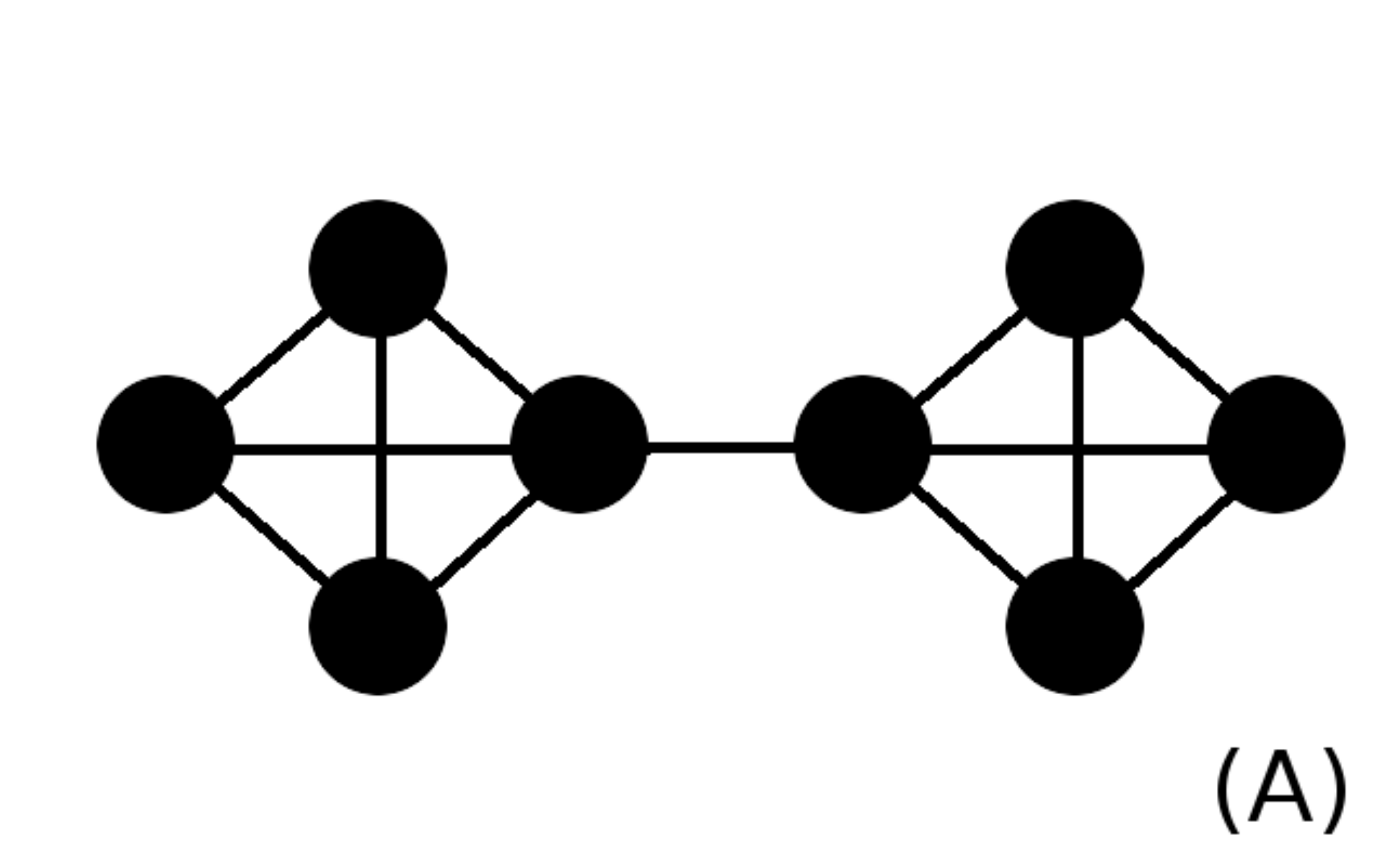}
%\hspace{2.3cm}
\hfill 
\includegraphics[width=1.7in]{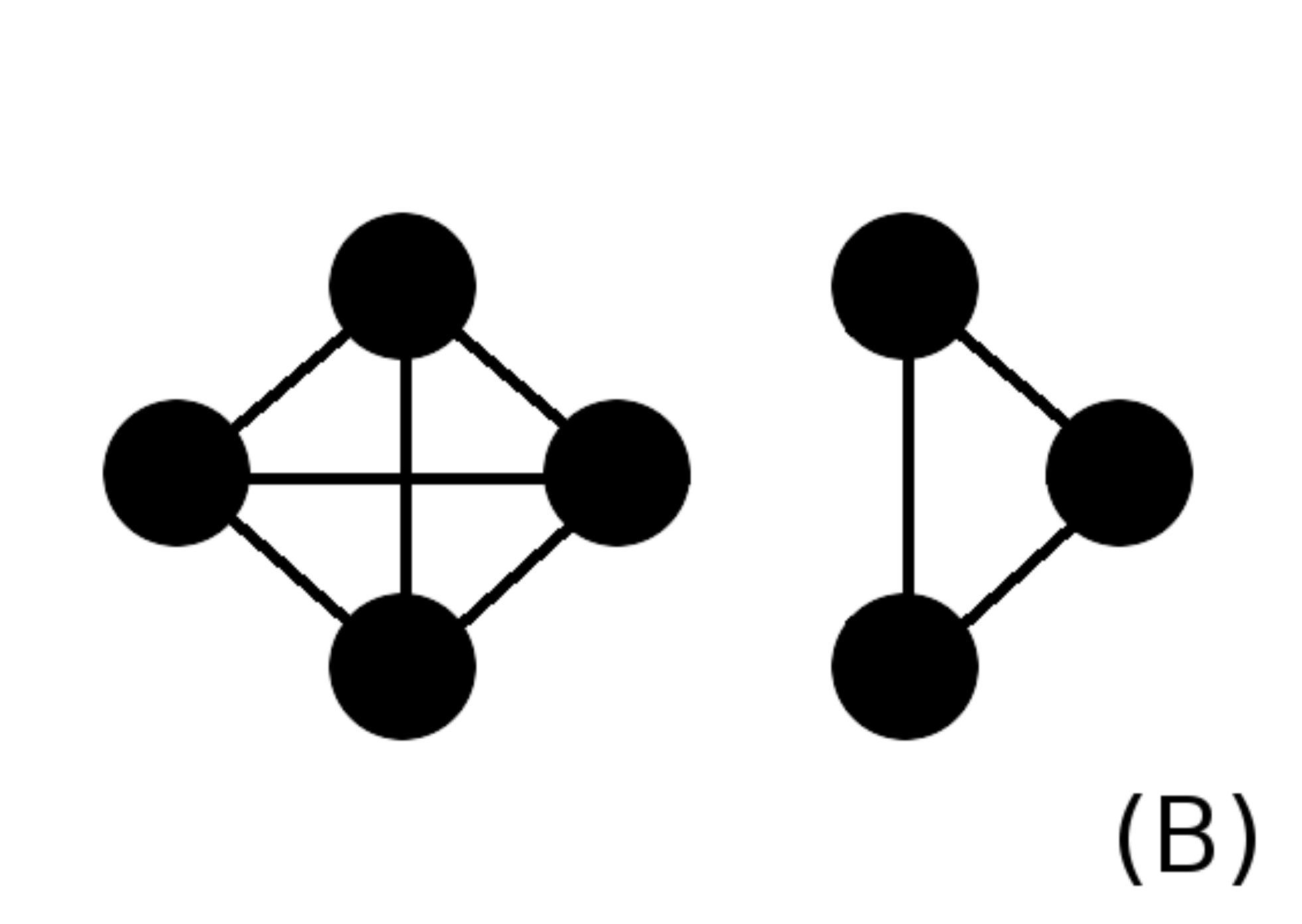}
\hspace{0.6cm}
}
\caption{Dynamics of a colony network where two all-to-all sub-networks are joined by a single connection: (A) network structure for a stable coexisting equilibrium; (B)  network structure after one of the two connecting colonies has collapsed (see text).}
\label{fig:network}
\end{figure}

We first consider two $N$-colony networks, both of which are connected all-to-all, and are joined together by a single connection between one colony in each network; the $N=4$ case is shown in Figure \ref{fig:network}A. In the symmetric case, in which $\gamma_i=\gamma_j$ for all $1\leq i,j\leq 2N$, the Jacobian matrix of the interior equilibrium point can be shown to have characteristic polynomial %\imp{Should we show the matrix form here?}

 $$\chi_{J_{NN}}(x)=\chi_{P}(x)\left[\chi_{P+NQ}(x)\right]^{2N-3}\chi_{P+\lambda_1Q}(x)\chi_{P+\lambda_2Q}(x),$$
 where $\lambda_1\geq \lambda_2$ are the roots of the quadratic equation $$\lambda^2-(N+2)\lambda+2=0.$$ 
The first two distinct factors are the same as those that appear in the characteristic polynomial (\ref{eq:charPoly}) of the $N$-nest all-to-all connected system. The latter two factors of $\chi_{J_{NN}}(x)$ are the result of the modified network structure, and correspond to the stability of the connection between the two colonies. Therefore the stability of the coexisting state once again reduces finding the eigenvalues of a $2\times 2$ matrix. %In the same way that we wrote the real part of the matrix $P+NQ$ as the function of $N$ $\sigma_N$ in Section \ref{sec:nNest}, we can write write the eigenvalues of $P+\lambda_1Q$ and $P+\lambda_2Q$ as the same function $\sigma_{\lambda_1}$ and $\sigma_{\lambda_2}$, respectively.
Just as in the all-to-all connected case above, for $\beta_u$ sufficiently small, all eigenvalues of $J_{NN}$ have negative real part, and the network is consequently stable. Because $\lambda_1\geq\lambda_2$, the eigenvalues of $P+\lambda_1Q$ are the first to pass through the imaginary axis as $\beta_u$ increases, thereby disrupting the overall stability of the network. Each eigenvalue corresponding to the stability of the separate two $N$-nest all-to-all connected colonies has negative real part, and conseqeuntly the instability must be in the connection between the two colonies. Thus, for sufficiently large $\beta_u$, one of the two colonies between which the two networks are connected will collapse, and the system will fragment into two independent networks, one of size $N$, the other of size $N-1$ (Figure \ref{fig:network}B). The dynamics within these two networks then become  those discussed in the beginning of Section \ref{sec:nNest}. %\imp{The line defined by the real part of the eigenvalues of $P+\lambda_1Q$ set equal to zero is generically to the right of the same line defined by the real part of the eigenvalues of $P+NQ$. Can show this using the fact that the real part is half the trace. This tells us that the part of the colony that connects the two sub-colonies loses stability first as $\beta_u$ increases, and so the colony will fragment into two separate colonies.}

\begin{figure}[H]
{\centering

\includegraphics[width=2.25in]{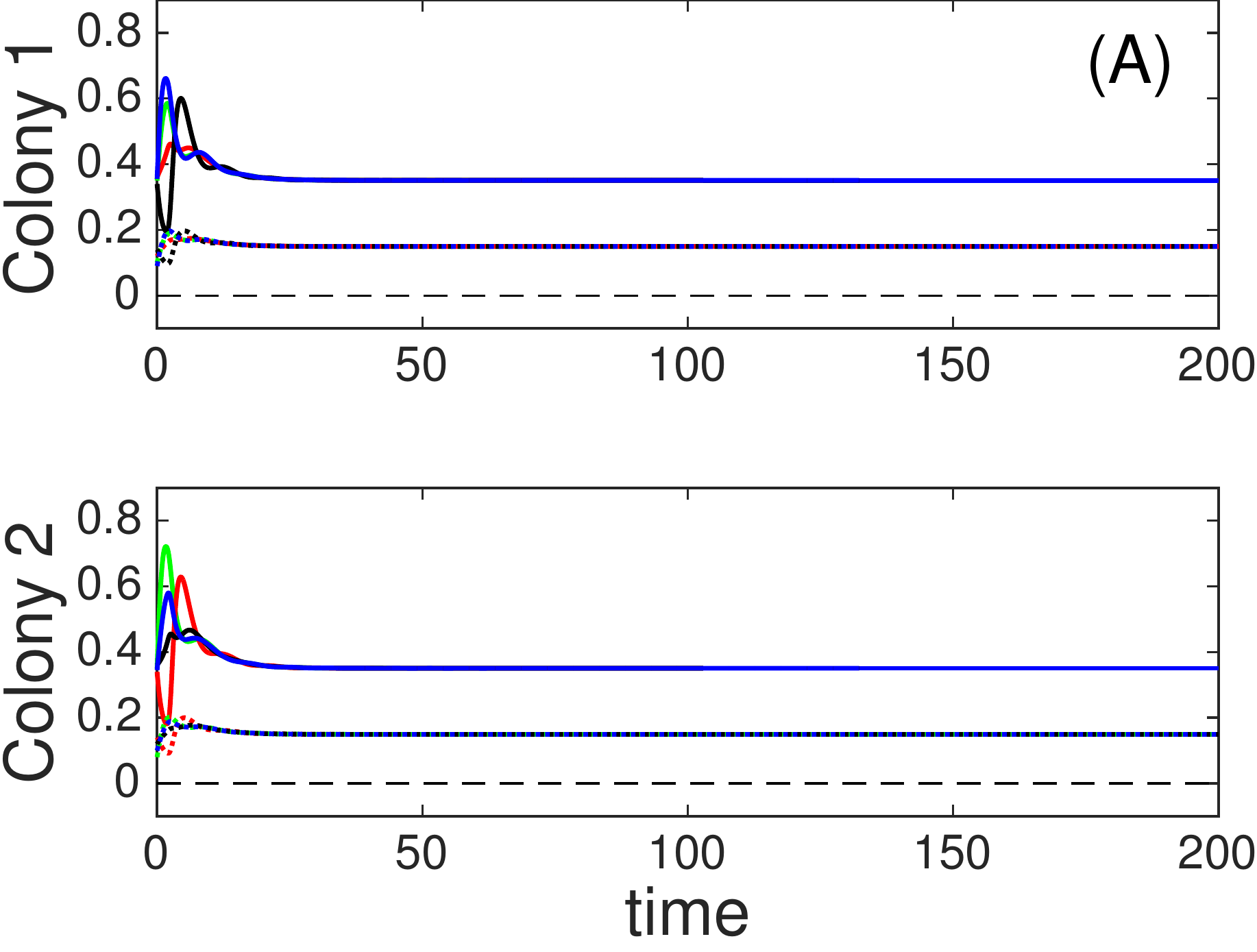}\hspace{5mm} \includegraphics[width=2.25in]{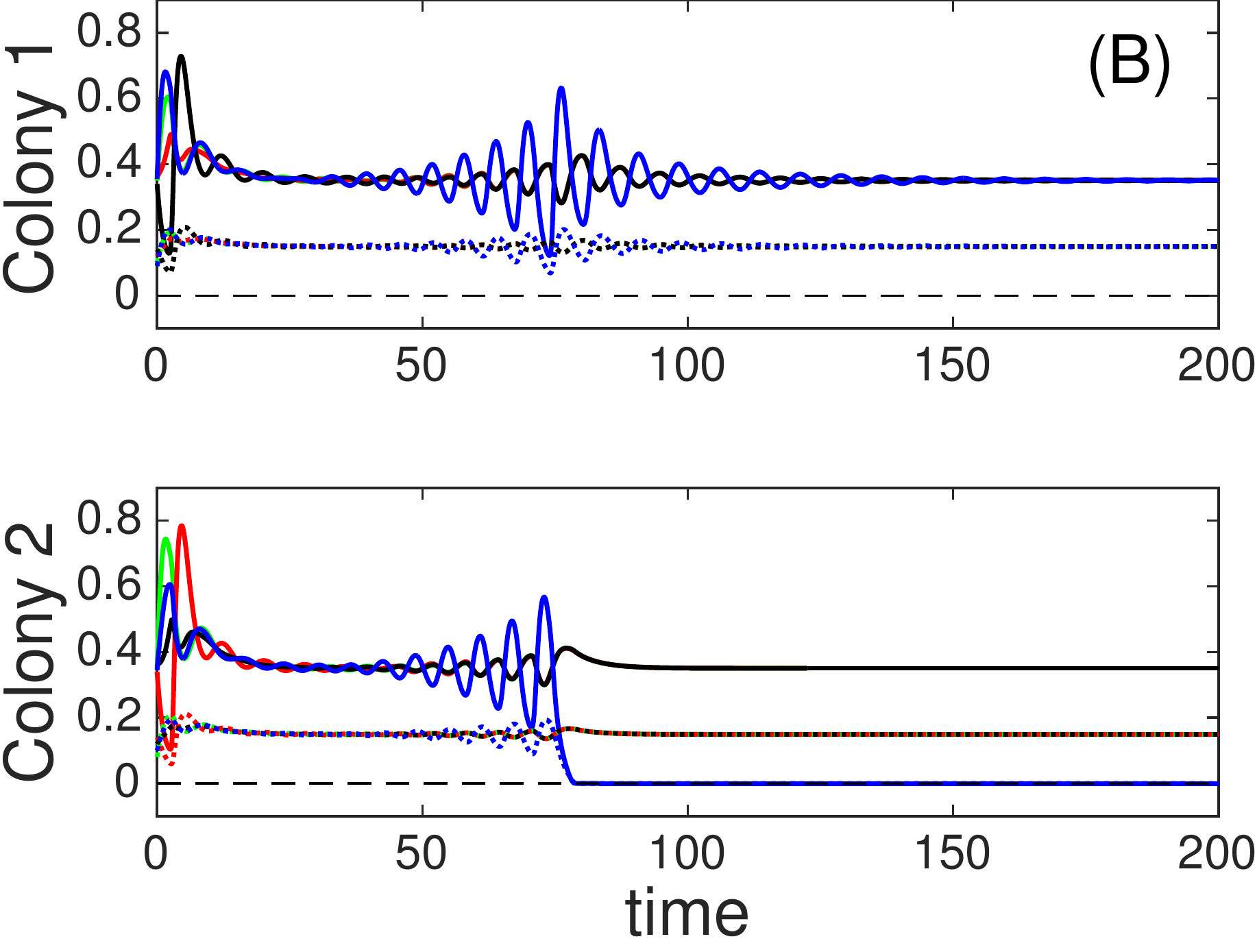}\\
\includegraphics[width=2.25in]{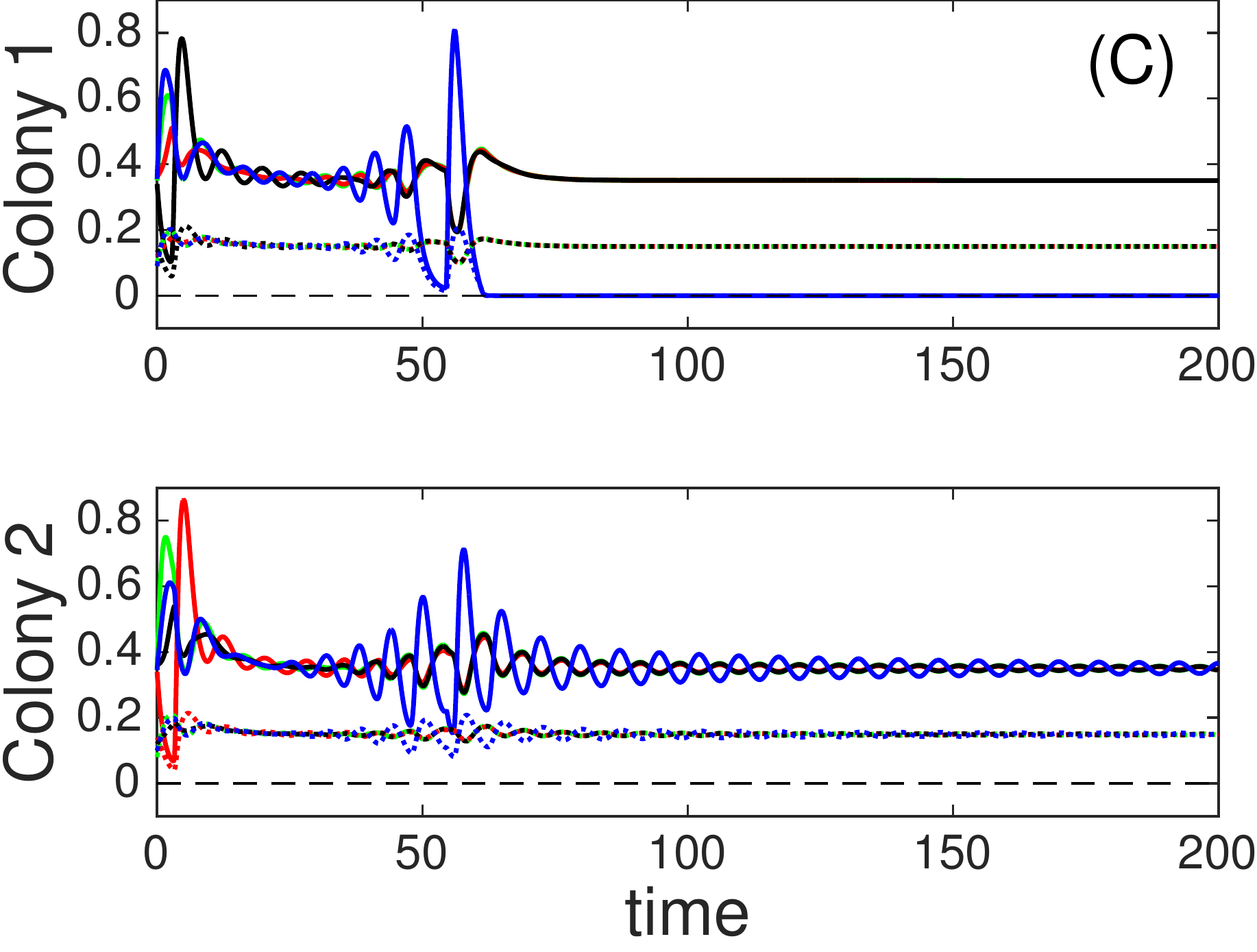}\hspace{5mm} \includegraphics[width=2.25in]{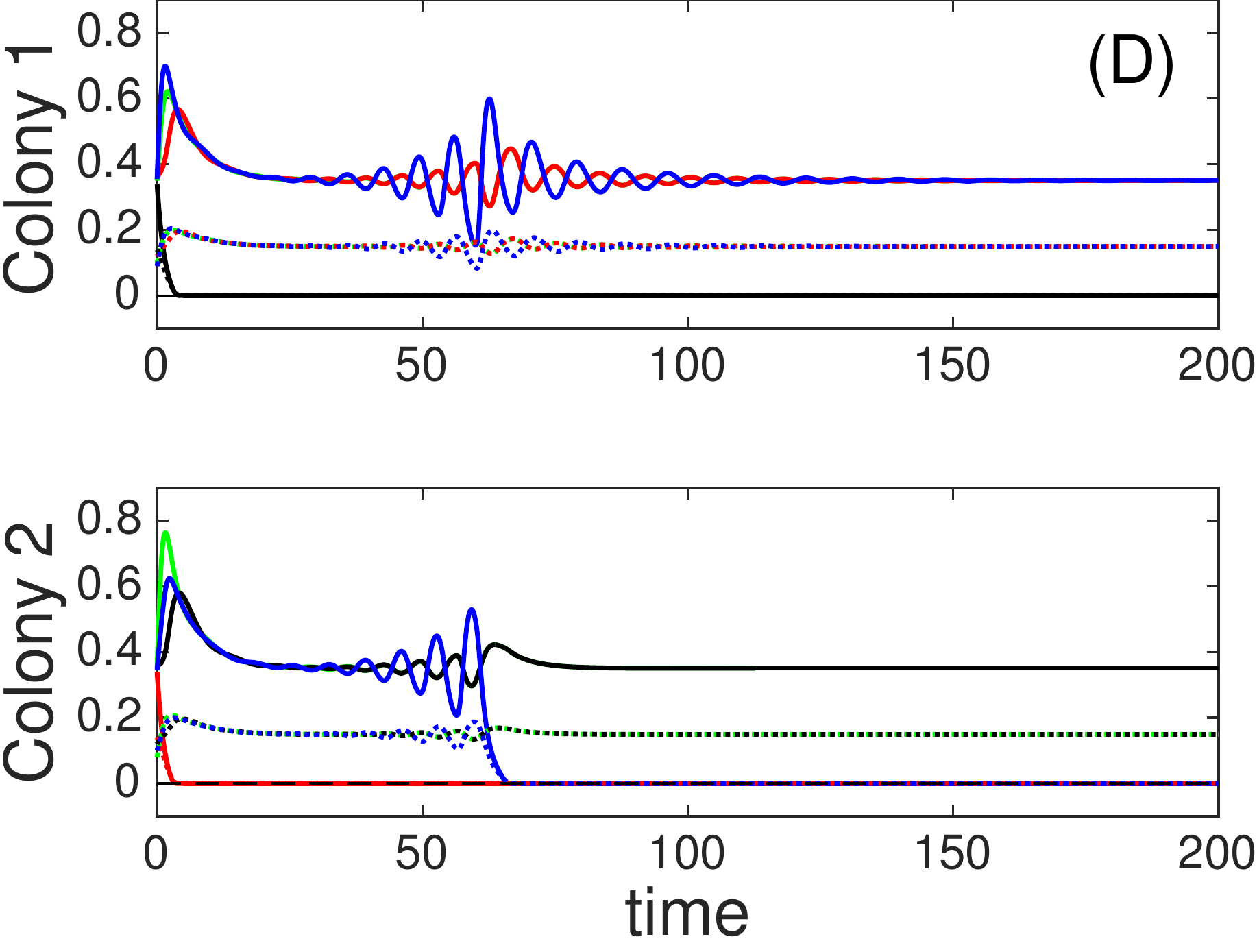}

}
\caption{Dynamics of two sub-networks of four colonies each, with $\beta_v=0.2$, joined via single connection . A. $\beta_u=0.55$. The system reaches a coexisting steady state in each nest. B. $\beta_u=0.63$. The coexisting steady state is unstable, and the connecting colony in the second sub-network collapses, resulting in two independent networks. C. $\beta_u=0.65$. The coexisting steady state is unstable, and the connecting colony in the first network collapses. D. $\beta_u=0.7$. One colony in the first sub-network collapses and two colonies in the second sub-network collapse, resulting in two independent networks, one comprised of three colonies, the other of two.}
\label{fig:2Col4nest}
\end{figure}

The case when $N=4$ is shown in Figure \ref{fig:2Col4nest}. For small $\beta_u$, the system reaches its coexisting steady state (Figure \ref{fig:2Col4nest}A), which undergoes a supercritical Hopf bifurcation as $\beta_u$ increases through 0.584, then loses stability entirely through a torus bifurcation at $\beta_u=0.610$. As predicted, after stability is lost, one of the two nests through which the two colonies are connected collapses in each case shown, resulting in two independent colonies (Figure \ref{fig:2Col4nest}B and C) comprised of at most three and four nests, respectively, though fewer nests are possible if $\beta_u$ is sufficiently large (Figure \ref{fig:2Col4nest}D). In Figure \ref{fig:2Col4nest}D, one of the colonies away from the connection in each network collapses before either of the connection-forming colonies does, resulting briefly in two networks of three colonies being connected by a single connection. The connection between the separate networks is still unstable, however, and one of the two connecting nests will collapse, resulting in two independent networks: one comprised of three colonies, the other of two. This is worse than the theory predicts, since the three-colony all-to-all connection network is still stable for $\beta_u=0.7$. The network structure therefore led to one additional colony collapsing: three total colonies collapse, compared to the two that theory predicts would collapse between two independent networks of four colonies each with the same migration rates. {In this way, the network structure negatively impacts the stability of the network.}

%%%%%%%%%%%%%%%%%%%%%%% 
\subsubsection{Ring of colonies}
 
\begin{figure}[!t] % H
{\centering

%\includegraphics[width=1.0in]{ring_network.eps}
%\hspace{2.4cm} 
%\includegraphics[width=1.0in]{line_network.eps}\
\includegraphics[width=4in]{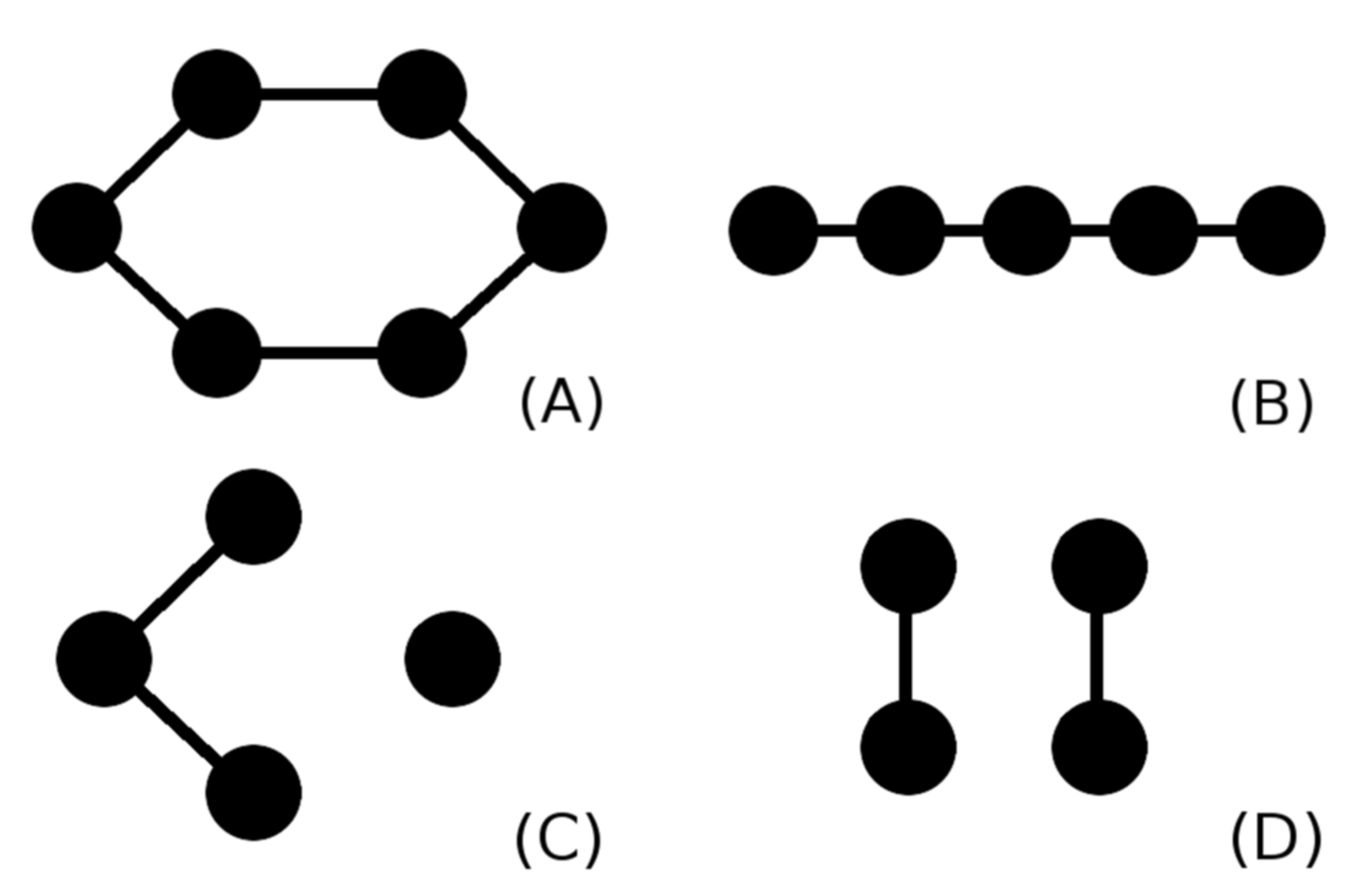}

}
\caption{Loss of stability for a ring of six nests. A. The network before stability is lost. B. The network after a single nest collapses. C and D. The network after two non-adjacent nests collapse. The network shown in each panel A-D corresponds to the final network structure in Figure \ref{fig:6nestRing} A-D, respectively.}
\label{fig:ringNetwork}
\end{figure}

As a second example of a tractable network configuration, we consider $N$ colonies connected in a ring. Much like each of the previously considered cases, a ring of nests can support a stable population of coexisting workers and cheaters for $\beta_u$ sufficiently small, and that stability is lost if $\beta_u$ becomes too large.
Determining the stability for such a ring is straightforward in the case when $N$ is even. Curiously, the coexistence state in a ring of any even size undergoes a Hopf bifurcation at the same critical value of $\beta_u$; specifically when $\sigma_N$ with $N=4$ passes through 0. The steady state then loses stability through a torus bifurcation, much like every other case considered thus far. In the case when $N$ is odd, the analytic form of the eigenvalues is too cumbersome to provide any insight, though numerical analysis suggests the coexistence steady state loses stability in a qualitatively identical manner as when $N$ even. After the ring loses stability, the network will break into a line of $N-1$ colonies (Figure \ref{fig:ringNetwork}B), and this line may then itself lose stability if $\beta_u$ is sufficiently large (Figure \ref{fig:ringNetwork}C and D).
 
% \imp{If $N$ is even, the Hopf bifurcation occurs as $\sigma_4$ passes through 0.}
 
%\imp{Include stability analysis of ring and line? Maybe in an appendix?}

\begin{figure}[H]
{\centering

\includegraphics[width=2.2in]{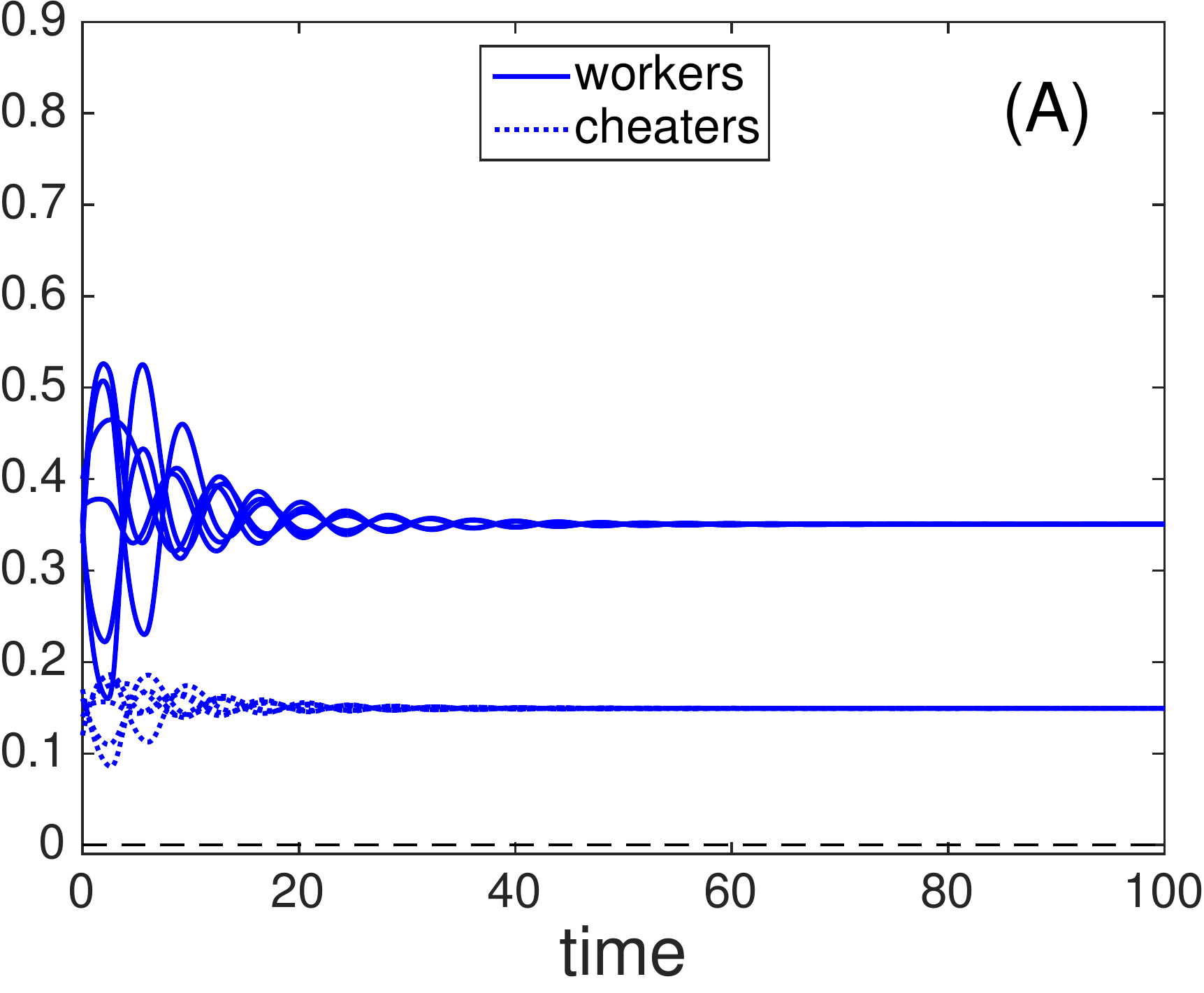}\hspace{8mm} \includegraphics[width=2.2in]{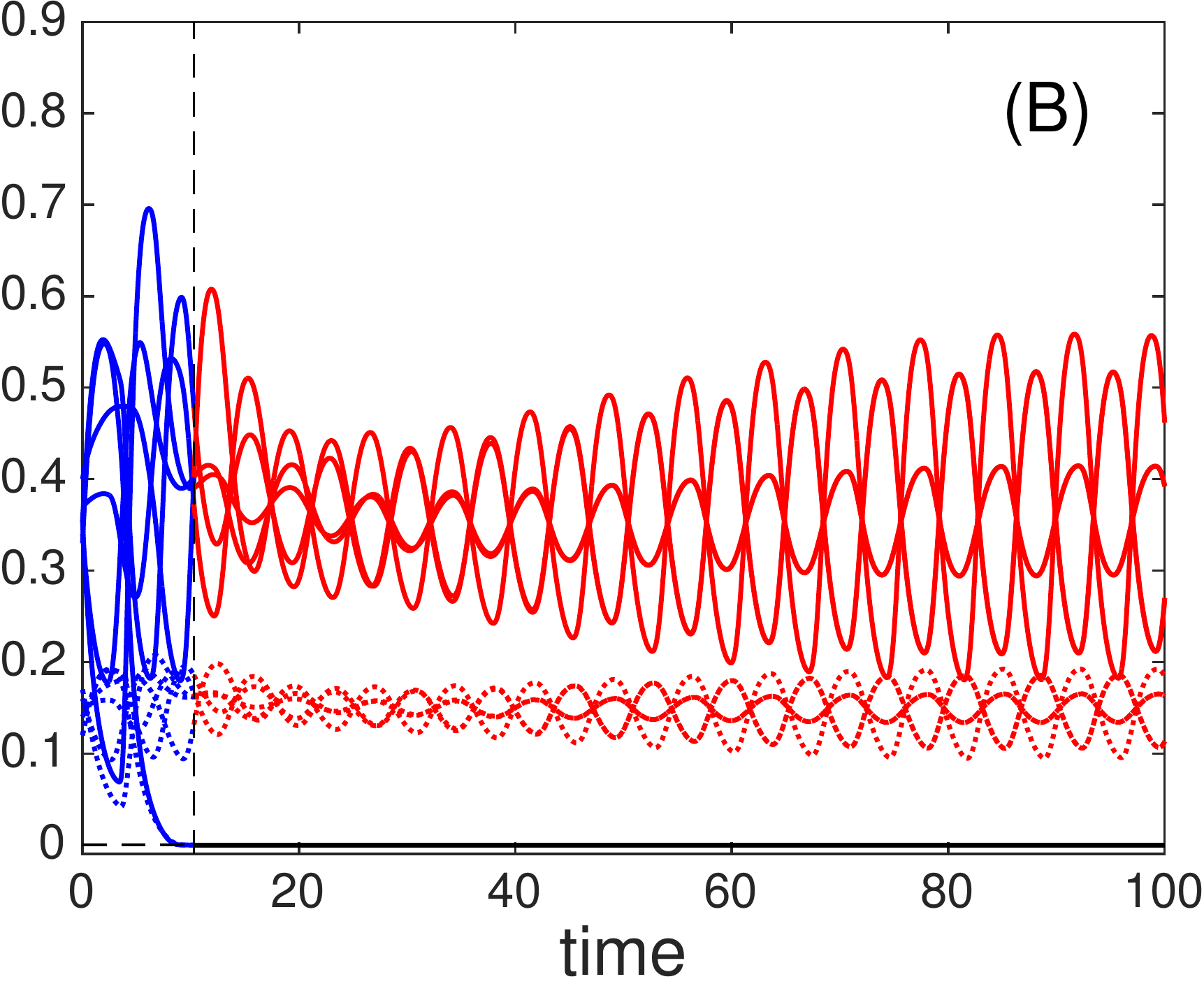}\\
\includegraphics[width=2.2in]{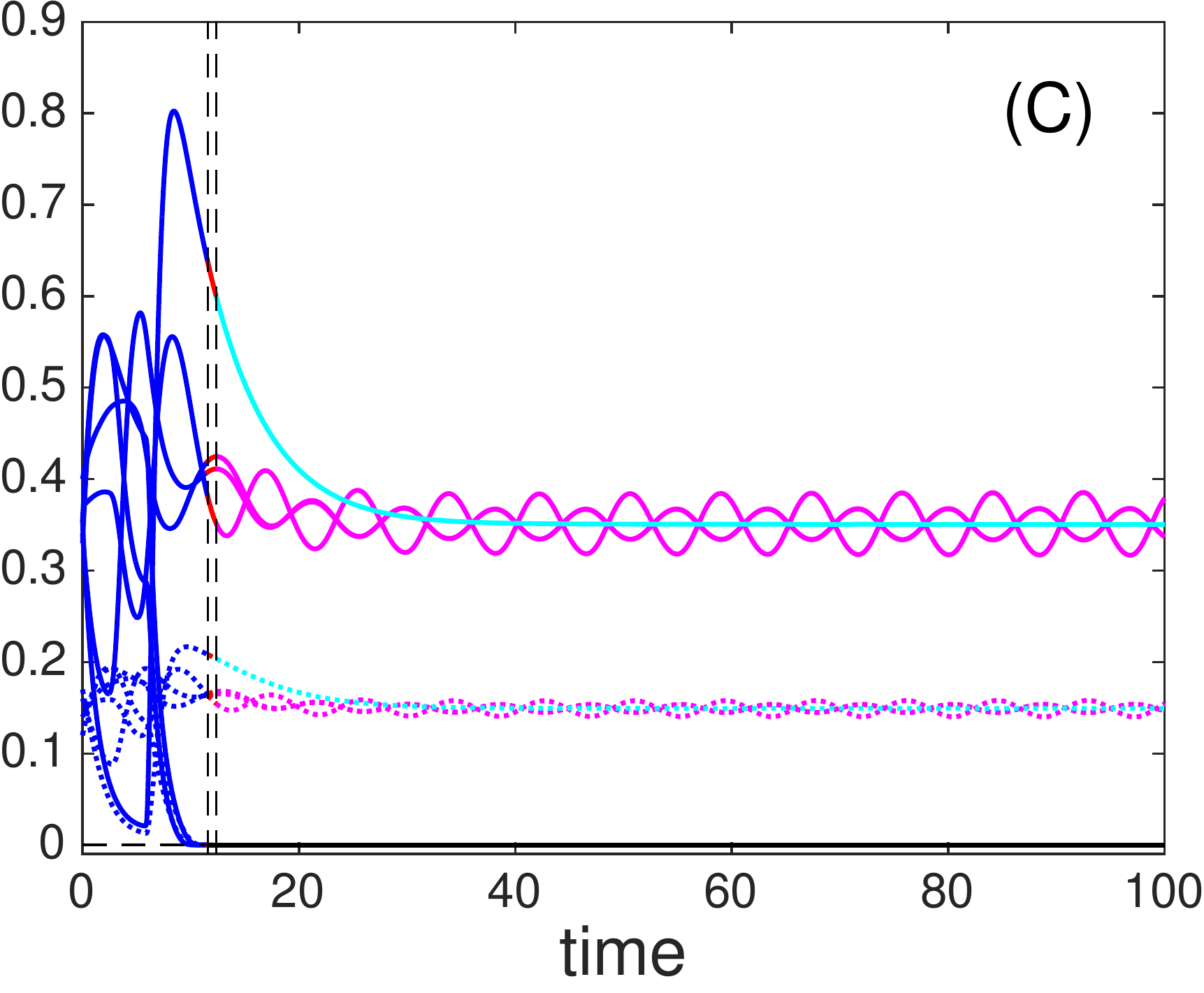}\hspace{8mm} \includegraphics[width=2.2in]{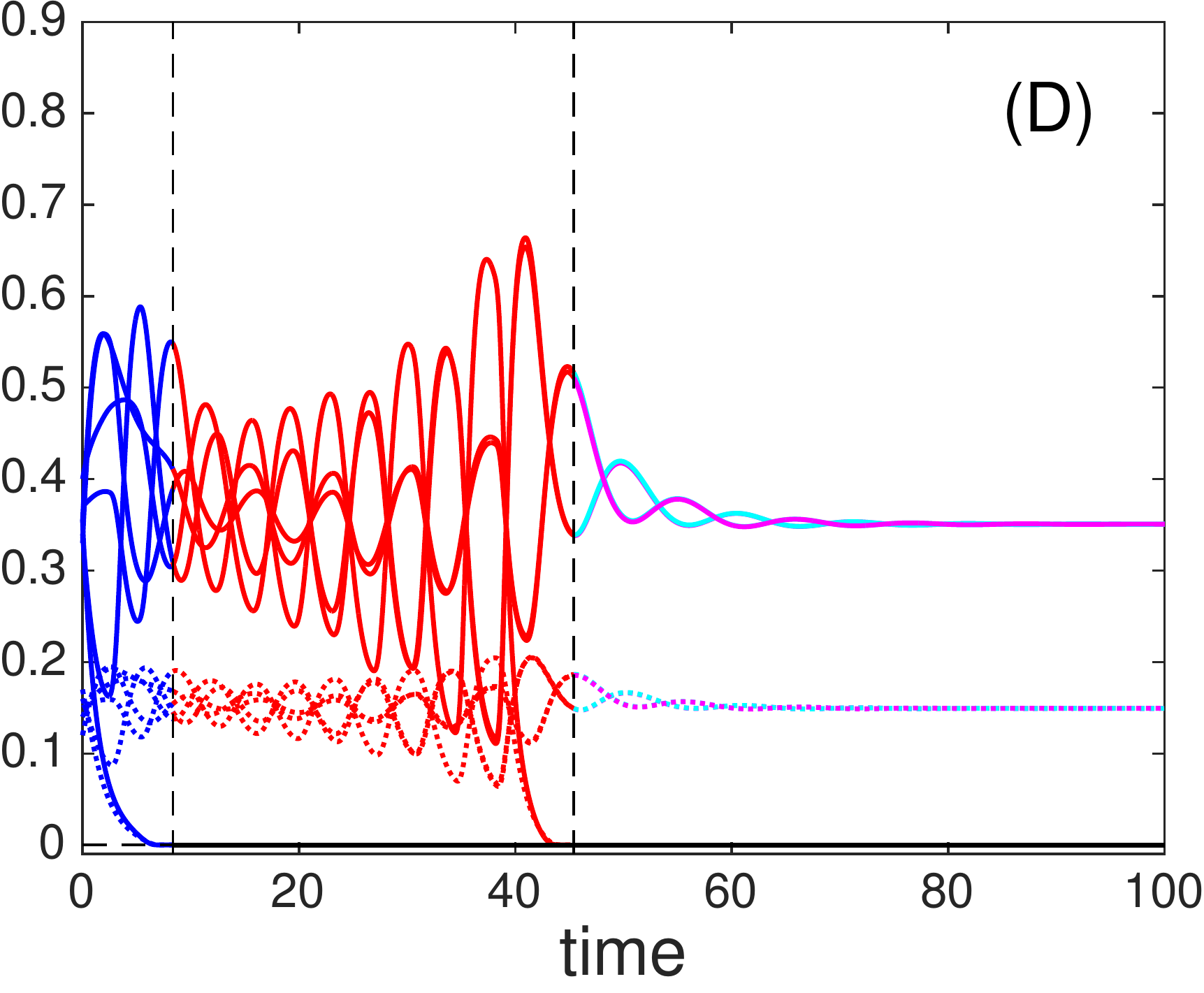}

}
\caption{Dynamics of six nests arranged in a ring, with $\beta_v=0.2$. Vertical dashed lines and color changes mark the collapse of a nest, and like-colored curves represent populations within nests that belong to the same network. A. $\beta_u=0.6$. The system reaches its coexisting steady state in each of the six nests. The network structure is shown in Figure \ref{fig:ringNetwork}A. B. $\beta_u=0.72$. A single nest collapses, resulting in a line of five nests oscillating over time, as shown in Figure \ref{fig:ringNetwork}B. C. $\beta_u=0.745$. The line of five nests loses stability entirely, and two nonadjacent nests collapse, resulting in two independent colonies: one comprised of three nests, the other of a single nest (Figure \ref{fig:ringNetwork}C). The colony consisting of three nests oscillates over time, while the single independent nest reaches a stable steady state. D. $\beta_u=0.75$. Two nonadjacent nests collapse, resulting in two independent colonies of two nests each (Figure \ref{fig:ringNetwork}D), both of which reach a coexisting steady state.}
\label{fig:6nestRing}
\end{figure}

Examples of the population dynamics for the case $N=6$ and $\beta_v=0.2$ are shown in Figure \ref{fig:6nestRing}. For sufficiently small $\beta_u$, the network remains intact and reaches its coexisting steady state (Figure \ref{fig:6nestRing}A). As the worker motility rate increases through $\beta_u=0.658,$ the coexisting equilibrium undergoes a supercritical Hopf bifurcation, and loses stability completely through a torus bifurcation as $\beta_u$ is increased through 0.710. The behavior of the system after stability is lost is highly dependent on both initial conditions and motility rates. Immediately after stability is lost in this case, a single nest will collapse, resulting in a line of five nests. This line-of-five-nests system approaches a stable limit cycle in Figure \ref{fig:6nestRing}B, then loses its stability through a torus bifurcation at $\beta_u=0.734$, after which another nest will necessarily collapse. Depending on which colony collapses, the network of five colonies can be reduced to a network of four colonies, and then to one network of three colonies and an independent, single colony (Figure \ref{fig:6nestRing}C), or to two independent networks of two colonies (Figure \ref{fig:6nestRing}D). In these various cases, the result is always that the initial network fragments into pieces. 

%%%%%%%%%%%%%%%%%
%%%%%%%%%%%%%%%%%

\begin{figure}[!t] % H
{\centering

\includegraphics[width=1.8in]{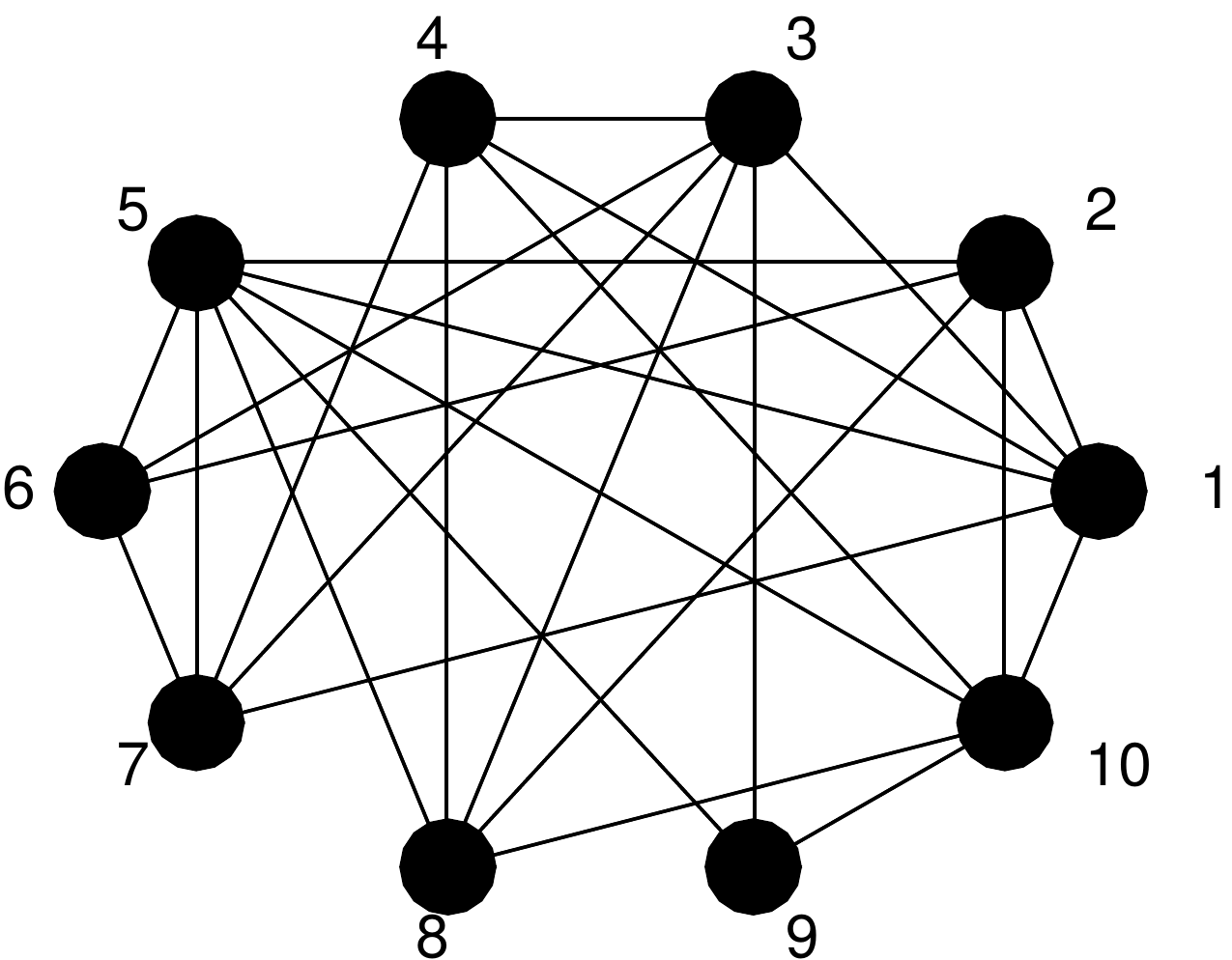}

\vspace{2mm}

\includegraphics[width=5in]{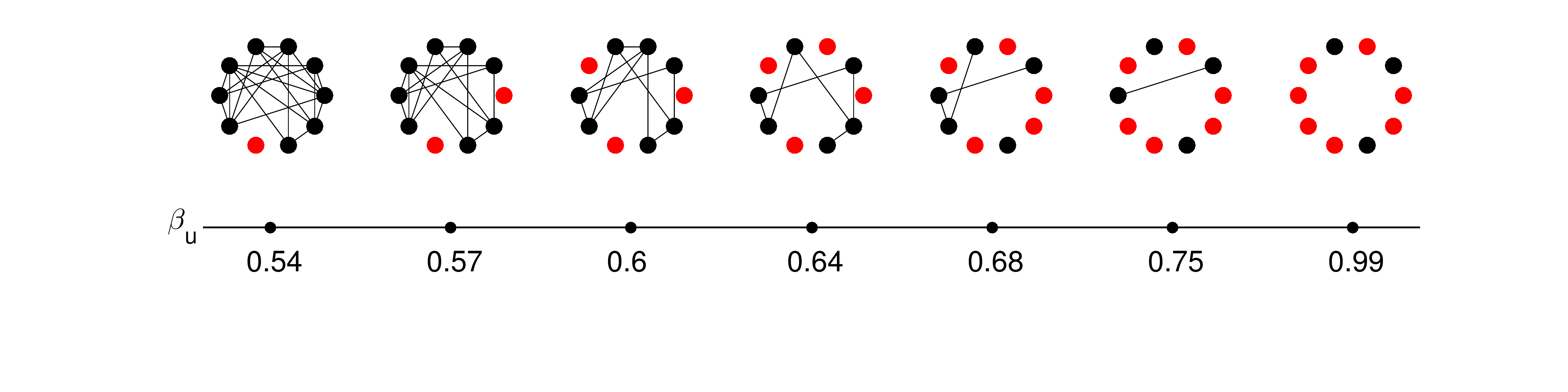}

}
\caption{A network of ten randomly connected colonies, using a probability of connection between each pair of $0.5$. The initial conditions were generated randomly as $u_i=0.35+u_\epsilon$ and $v_i=0.15+v_\epsilon$, where $u_\epsilon$ was selected at random from the interval $[-0.1,0.1]$ and $v_\epsilon$ from $[-0.06,0.06]$. For small $\beta_u$, the colony network is stable, but as the parameter is increased, the network loses stability. The figure shows the observed value of $\beta_u$ at which a colony collapses, and the resulting configuration of the network. Black circles denote active colonies, and red circles denote collapsed colonies.}
\label{fig:10nests}
\end{figure}

These results, along with the analysis of two colonies joined via a single connection in Section \ref{sec:twoColonies}, suggest that the size of an interconnected colony is limited by the motility rate of the workers, and independent of colony network structure.  To further study this conjecture, we consider a network of ten nests connected at random (nest $i$ connected to nest $j$ with probability $0.5$), and determine the stability of the network at varying motility rates. Figure \ref{fig:10nests} illustrates how the colony breaks apart through nest collapse as $\beta_u$ is increased. The upper network is the original, randomly connected starting condition in this case study. The seven lower networks show the progression of nest collapse over increased $\beta_u$. As expected, for sufficiently small $\beta_u$, the ten-nest as is stable, but as the rate is increased past a critical threshold, one of the ten nests collapses, leaving a nine-nest colony. This continues as $\beta_u$ is increased until the colony collapses into three independent nests. While we only present results from a single random realization, the qualitative behavior seems to be consistent over repeated simulations (data not shown). Statistical analysis of network-breakdown over a large number of simulations could prove useful in understanding the role of motility in colony network collapse, which we leave as an open question.

%%%%%%%%%%%%%%%%%%%%%%%%%%%%% 
%%%%%%%%%%%%%%%%%%%%%%%%%%%%% 

 \section{Discussion}
 
We have introduced a model combining within-colony population dynamics and between-colony migratory dynamics of a social organism that reproduces asexually. The population dynamics are rooted in a modified ecological public goods game, not unlike those found in \cite{Wakano}, in which workers provide a public good that is freely available to every individual within the colony, while free-riding cheaters do not produce any public good. The migratory dynamics are based on the assumption that the populations will always move toward a colony of higher fitness than their current colony, or equivalently, away from colonies under stress \cite{deForest2}. Analysis of our model indicates that, while cheaters can greatly reduce the overall fitness of any nest they invade, their ability to migrate can act as a stabilizing factor in a network of nests. In particular, if cheaters are sufficiently fast relative to workers, the likelihood that any given nest will collapse is reduced, thereby stabilizing the network. Importantly, our results suggest that the cheaters must migrate at a sufficiently high rate \emph{relative to the workers}. If the workers are significantly faster than the cheaters, they will constantly be moving away from the lowered fitness caused by  cheaters, not allowing themselves time to establish a stable population within a nest. On the other hand, if the cheaters are sufficiently fast, they spread and occupy each available nest, removing any higher-fitness options from the workers. The workers are thereby forced to stay in the cheater-occupied nest, having no better options to attract them. {In spatial ecology, this distribution in which individuals cannot increase their fitness by moving to a different nest is called the ideal free distribution \cite{Cantrell,cosner2005,Cressman}.}

Two examples of organisms whose social structures generally meet the assumptions in our model are the queenless ant species {\it Pristomyrmex punctatus} and the Cape honeybee {\it Apis mellifera capensis}. Populations of {\it P. punctatus} found in central Japan maintain a subpopulation of genetic cheaters that contribute nothing to the public good, in contrast with the productive workers, who forage outside of the nest for food and tend to the young within the nest \cite{Dobata1,Dobata2}. The cheaters naturally add stress to the nest and workers in the form of increased demand on public efforts, though colony collapse due to these cheaters has not been observed \cite{Dobata1}. Not surprisingly, these cheaters spread between colonies, taking advantage of the free public goods while harming the overall fitness of each colony \cite{Dobata1,Tsuji}. Of particular significance is the estimation that cheaters tend to migrate between nests at a rate roughly three orders of magnitude faster than their worker counterparts \cite{Dobata2,Tsuji}. In \cite{Dobata1}, the authors interpret this motility difference to indicate that cheaters might be able to abandon a nest before it collapses due to their own presence.  {Our results agree with this hypothesis, and further suggest that if the workers migrated at a rate similar to the cheaters, then colonies would face collapse.}

Cape honeybee workers behave as parasites themselves, invading nearby honeybee colonies and acting as pseudoqueens, ultimately resulting in the loss of the colony's true queen and the collapse of the colony \cite{Neumann}. While no estimates of Cape honeybee and their hosts' migration rates seem to exist, it is certainly reasonable to expect that these rates are faster than those of the previously discussed ants, as the bees fly instead of walk, are not impeded by geography, etc. If so, the respective behaviors of these ants and bees agree qualitatively with our model's prediction: the cheaters in {\it P. punctatus} colonies migrate quickly relative to workers, and the populations manages to avoid colony collapse, while our model would predict that the native worker honeybees migrate at a rate closer to that of the cheaters, resulting in the observed colony collapse.

%workers reproduce, care for broods, and forage outside of the nest, while genetic cheaters only reproduce. One proposed mechanism by which these ants might avoid colony collapse is their ability to migrate. Cheaters ants tend to migrate at a much faster rate than workers, an advantage that could allow them to avoid their own extinction by emigrating from a failing nest. These results agree with the conjecture offered in \cite{Dobata2}: the cheaters use fast migration as a mechanism to abandon a failing colony, thereby avoiding collapse while spreading to invade other colonies. 

%The ant species \emph{Pristomyrmex punctatus} is an example of such an organism. Workers of this species reproduce, care for broods, and forage outside of the nest, while genetic cheaters only reproduce. One proposed mechanism by which these ants might avoid colony collapse is their ability to migrate. Cheaters ants tend to migrate at a much faster rate than workers, an advantage that could allow them to avoid their own extinction by emigrating from a failing nest. 

Our analysis predicts that one or more nests will collapse if the workers are too quick to migrate away from stress induced by cheaters; however, the particular nests that collapse are highly sensitive to changes in motility rates and initial conditions (Figures \ref{fig:betaBasin} and \ref{fig:basinSym}). This suggests that the dynamics of the system might be changed significantly if we include in our model a small stochastically varying migration term between each pair of connected nests. Indeed, preliminary simulations suggest that allowing such random migration in the worker population can stabilize the colony, preventing nest collapse in an otherwise unstable parameter regime. Random migration in the cheater population does not affect the qualitative dynamics, however. This suggests an interesting trade off between two types of migration in the worker population: the faster the workers move away from fitness-reducing stress, the more likely a nest will collapse; however, faster random migration between nests can help stabilize the colony. We leave this observation as a comment, and any questions about the effects of random motility open. Such stabilization through stochastic processes has been studied in, e.g., \cite{Arnold}. 

 {We mention in Section \ref{sec:twoNest} that at least one nest must survive the collapse of a network of nests. This is because the motility function in our model assumes that individuals move only toward nests of higher fitness, and consequently individuals cannot migrate if only a single nest remains.  Future developments of this model should therefore include between-nest dynamics, allowing individuals to leave a nest without immediately entering another nest. This addition would also allow one to consider the potentially important biological factor of the cost of dispersal \cite{Bonte}.}  Additionally, future work should include the possibility of a more dynamic network, as the merging of colonies has been observed in {\it P. punctatus} \cite{satow}. While we contend that these omissions do not critically affect our results, their inclusion could be useful in understanding related aspects of migration, including optimal migration rates.

% \begin{itemize}
% \item Past modeling efforts
%\begin{itemize}
%\item bee nest selection
%\item IFD
%\item public goods games
%\end{itemize}
% \item the possibility of noise stabilizing a colony after the torus bifurcation
% \item the counterintuitive result that motility hurts workers
% \item though our results suggest that one or more nests will collapse, predicting which ones will is likely impossible
% \item introducing the possibility of genetic heterogeneity could lead to more interesting results
% \end{itemize}
 
\section{Acknowledgements}

This work was supported by National Science Foundation Grant CMMI-1463482. We would like to thank Russ deForest for discussions and helpful comments.

\section{Appendix}
 \subsection{Slowly decaying public good}\label{sec:epsilon}
 
 If we consider the public good available within a nest to be a dynamic variable, say $\phi$, then the within-nest dynamics of the workers and cheaters can take the form
 \begin{equation*}
\begin{aligned}
\dot u&=u\left[\tilde r_u\phi-c-\gamma(u+v)-\mu_u\right]\\
\dot v &= v\left[\tilde r_v\phi-\gamma(u+v)-\mu_v\right]\\
\dot \phi &= c u-\left[b(u+v)+d\right]\phi,
\end{aligned}
\end{equation*}
 where $a$ is the production rate of the public good, $b$ is the consumption rate by both the workers and cheaters, and $d$ is a natural decay rate. This decay rate could biologically correspond to spoilage of food, for instance. If we assume the public good reaches its steady state much more quickly than the workers and cheaters, then we can replace $\phi$ with its steady state in the $u$ and $v$ equations:
  \begin{equation*}
\begin{aligned}
\dot u&=u\left[\tilde r_u\frac{cu}{b(u+v)+d}-c-\gamma(u+v)-\mu_u\right]\\
\dot v &= v\left[\tilde r_v\frac{cu}{b(u+v)+d}-\gamma(u+v)-\mu_v\right].
\end{aligned}
\end{equation*}
 Defining $r_u=\tilde r_u/b$, $r_v=\tilde r_v/b$, and $\epsilon=d/b$, we obtain system (\ref{eq:noMigEps}), and $0<\epsilon<<1$ as long as $0<d<<b$.

% DRAFT - COMMENTED OUT
 
% \subsection{Stability of two connected colonies}
 
% \[
%J_{NN}=\begin{bmatrix}
%    F+(N-1)H & -H & -H & \dots  & -H & 0 & 0 & 0 & \dots & 0 \\
%    -H & F+(N-1)H & -H & \dots  & -H & 0 & 0 & 0 & \dots & 0 \\
%    \vdots & \vdots & \vdots & \ddots & \vdots &  \vdots & \vdots & \vdots & \ddots & \vdots \\
%    -H & -H & -H & \dots  & F+(N-1)H & -H & 0 & 0  & \dots & 0\\
%    0 & \dots & 0 & 0 & -H & F+(N-1)H & -H & -H & \dots  & -H  \\
%    0 & \dots & 0 & 0 & 0 & -H & F+(N-1)H & -H & \dots  & -H  \\
%    \vdots & \vdots & \vdots & \ddots & \vdots  \\
%     0 & \dots & 0 & 0 & 0 &-H & -H & -H & \dots  & F+(N-1)H
%\end{bmatrix},
%\]

% DRAFT - COMMENTED OUT

% \[
%J_{NN}=\begin{bmatrix}
%    J_N & -H_{ll} \\
%    -H_{ur} & J_N
%\end{bmatrix},
%\]


\section{References}

 
\begin{thebibliography}{45}

%%%%%
\bibitem{Archetti1} 
Archetti, M., Ferraro, D.A. and Christofori, G., 2015. Heterogeneity for IGF-II production maintained by public goods dynamics in neuroendocrine pancreatic cancer. {\it Proc. Natl. Acad. Sci.}, 112(6), 1833-1838.
Vancouver	

%Cancer public goods dilemma

%%%%%%%%%%
 \bibitem{Archetti2} Archetti, M. and Scheuring, I., 2012. Game theory of public goods in one-shot social dilemmas without assortment. {\it J. Theor. Biol.}, 299, 9-20. %public goods review

%%%%%%%%%
 \bibitem{Arnold} Arnold, L., 1990. Stabilization by noise revisited. {\it ZAMM}, 70(7), 235-246. %white noise stabilization
 
 %%%%%%%%
 \bibitem{Bonte} Bonte, D., Van Dyck, H., Bullock, J.M., Coulon, A., Delgado, M., Gibbs, M., Lehouck, V., Matthysen, E., Mustin, K., Saastamoinen, M. and Schtickzelle, N., 2012. Costs of dispersal. {\it Biol. Rev.}, 87(2), 290-312. %cost of dispersal
 
 %%%%%
 \bibitem{Cantrell} Cantrell, R.S., Cosner, C. and Lou, Y., 2012. Evolutionary stability of ideal free dispersal strategies in patchy environments. {\it J. Math. Biol.}, 65(5), pp.943-965.
 
 
 %%%%%
 \bibitem{Celiker}  Celiker, H. and Gore, J., 2012. Competition between species can stabilize public-goods cooperation within a species. {\it Molecular Systems Biology}, 8(1), p.621.
 
%%%%%
\bibitem{cosner2005} Cosner, C., 2005. A dynamic model for the ideal-free distribution as a partial differential equation. 
\textit{Theor. Popul. Biol.} {\bf 67} 101-108.

 %%%%
 \bibitem{Cressman} Cressman R, K\u{r}ivan V., 2006. Migration dynamics for the ideal free distribution. {\it Am. Naturalist}, 168(3), pp.384-397.
 
 \bibitem{Damore} Damore, J.A. and Gore, J., 2012. Understanding microbial cooperation. {\it Journal of theoretical biology}, 299, pp.31-41.
 
 %%%%%%%%%%%%
  \bibitem{deForest1} deForest, R., Belmonte, A., 2013. Spatial pattern dynamics due to the fitness gradient flux in evolutionary games. {\it Phys. Rev. E.}, 87(6), 062138.
  
  %%%%%%%%%%%%%%%%%%%%% TPB format line %%%%%%%%%%%%%
  
  %%%%%%%%
 \bibitem{deForest2} deForest, R., Belmonte, A. 2015 A game-theoretic dispersal mechanism for interacting populations. {\it Preprint.}
 
 %%%%%%%
\bibitem{Dobata1} Dobata, S., Sasaki, T., Mori, H., Hasegawa, E., Shimada, M. and Tsuji, K., 2011. Persistence of the single lineage of transmissible ?social cancer?in an asexual ant. {\it Molecular ecology}, 20(3), pp.441-455.
%discusses impact of migration on difference between cape honey bee and ant

%%%%%%%%
 \bibitem{Dobata2} Dobata, S. and Tsuji, K., 2013. Public goods dilemma in asexual ant societies. {\it Proc Natl Acad Sci}, 110(40), pp.16056-16060.


%%%%%
\bibitem{dudgeon2017} 
Dudgeon, C.L., Coulton, L., Bone, R., Ovenden, J.R. and Thomas, S., 2017. Switch from sexual to parthenogenetic reproduction in a zebra shark. {\it Scientific reports}, 7, p.40537.

%%%%%%%
\bibitem{Fehr} Fehr, E. and G\"achter, S., 2000. Cooperation and punishment in public goods experiments. {\it The American economic review}, 90(4), pp.980-994.

%%%%%%%
\bibitem{Gerlee} Gerlee, P. and Altrock, P.M., 2017. Extinction rates in tumour public goods games. {\it Journal of The Royal Society Interface}, 14(134), p.20170342.

%%%%%%%%%
% \bibitem{Greig} Greig D, Travisano M. The Prisoner's Dilemma and polymorphism in yeast SUC genes. Proceedings of the Royal Society of London B: Biological Sciences. 2004 Feb 7;271(Suppl 3):S25-6. %prisoner's dilemma yeast
 
 
 %%%%%
 \bibitem{Harris} Harris, J. and Ermentrout, B., 2015. Bifurcations in the Wilson--Cowan equations with nonsmooth firing rate. {\it SIAM Journal on Applied Dynamical Systems}, 14(1), pp.43-72.

%%%%%
\bibitem{Hauert} 
Hauert, C., Wakano, J.Y. and Doebeli, M., 2008. Ecological public goods games: cooperation and bifurcation. {\it Theoretical population biology}, 73(2), pp.257-263.

%%%%%
\bibitem{Heppner} 
Heppner, F. and Grenander, U., 1990. A stochastic nonlinear model for coordinated bird flocks. {\it The ubiquity of chaos}, pp.233-238.
%bird flocking model

%%%%%
\bibitem{Leine} Leine, R.I. and Nijmeijer, H., 2004. Bifurcations of equilibria in non-smooth continuous systems. In {\it Dynamics and Bifurcations of Non-Smooth Mechanical Systems} (pp. 125-176). Springer Berlin Heidelberg.

%%%%%
\bibitem{levin2014} 
Levin, S.A., 2014. Public goods in relation to competition, cooperation, and spite. {\it Proc Natl Acad Sci}, 111(Supplement 3), pp.10838-10845.


%%%%%
\bibitem{Martin} 
Martin, S.J., Beekman, M., Wossler, T.C. and Ratnieks, F.L., 2002. Parasitic Cape honeybee workers, Apis mellifera capensis, evade policing. {\it Nature}, 415(6868), pp.163-165.
%cape honey bee


%%%%%
\bibitem{mittwoch1978} 
Mittwoch, U., 1978. Parthenogenesis. {\it Journal of Medical Genetics}, 15(3), p.165.

%%%%%
\bibitem{Neumann}
Neumann, P. and Moritz, R., 2002. The Cape honeybee phenomenon: the sympatric evolution of a social parasite in real time? {\it Behavioral Ecology and Sociobiology}, 52(4), pp.271-281.

%%%%%
\bibitem{Oakland} Oakland, W.H., 1987. Theory of public goods. {\it Handbook of public economics, 2}, pp.485-535.

%%%%%
 \bibitem{Partridge} Partridge, B.L., 1982. The structure and function of fish schools. {\it Scientific american}, 246(6), pp.114-123. %fish motility survival

%%%%%
\bibitem{Perc} Perc M,, G\'omez-Garde\~nes J,, Szolnoki A,, Flor\'ia LM,, Moreno Y., 2013. Evolutionary dynamics of group interactions on structured populations: a review. {\it Journal of the royal society interface}, 10(80), p.20120997.

%%%%%
\bibitem{Reina} Reina, A., Marshall, J.A., Trianni, V. and Bose, T., 2017. Model of the best-of-N nest-site selection process in honeybees. {\it Physical Review E}, 95(5), p.052411.
%honeybee nest selection

%%%%%
\bibitem{Riehl} Riehl, C. and Frederickson, M.E., 2016. Cheating and punishment in cooperative animal societies. {\it Phil. Trans. R. Soc. B}, 371(1687), p.20150090.

%%%%%
\bibitem{satow} 
Satow, S., Satoh, T. and Hirota, T., 2013. Colony fusion in a parthenogenetic ant, Pristomyrmex punctatus. {\it Journal of insect science}, 13(1), p.38.

%%%%%
\bibitem{Shigesada} 
Shigesada, N., Kawasaki, K. and Teramoto, E., 1979. Spatial segregation of interacting species. {\it Journal of Theoretical Biology}, 79(1), pp.83-99.

%%%%%
\bibitem{simon2002} 
Simon, J.C., Rispe, C. and Sunnucks, P., 2002. Ecology and evolution of sex in aphids. {\it Trends in Ecology \& Evolution}, 17(1), pp.34-39.

%%%%%%
\bibitem{Szolnoki} Szolnoki, A. and Perc, M., 2010. Reward and cooperation in the spatial public goods game. {\it EPL (Europhysics Letters)}, 92(3), p.38003.

%%%%%
 \bibitem{Thompson} Thompson, W.A., Vertinsky, I. and Krebs, J.R., 1974. The survival value of flocking in birds: a simulation model. {\it The Journal of Animal Ecology}, pp.785-820.%birds flocking survival

%%%%%
 \bibitem{Tsuji} Tsuji, K. and Dobata, S., 2011. Social cancer and the biology of the clonal ant Pristomyrmex punctatus (Hymenoptera: Formicidae). {\it Myrmecol News}, 15, pp.91-99.
 
 %%%%%%
 \bibitem{Turnbull} Turnbull, G.A., Morgan, J.A.W., Whipps, J.M. and Saunders, J.R., 2001. The role of bacterial motility in the survival and spread of Pseudomonas fluorescens in soil and in the attachment and colonisation of wheat roots. {\it FEMS Microbiology Ecology}, 36(1), pp.21-31.%bacterial motility survival
 
 %%%%%
 \bibitem{Tyson} Tyson, R., Lubkin, S.R. and Murray, J.D., 1999. A minimal mechanism for bacterial pattern formation. {\it Proceedings of the Royal Society of London B: Biological Sciences}, 266(1416), pp.299-304.%bacterial pattern formation

%%%%%
 \bibitem{Velicer} Velicer, G.J., Kroos, L. and Lenski, R.E., 2000. Developmental cheating in the social bacterium Myxococcus xanthus. {\it Nature}, 404(6778), pp.598-601. %cheater bacteria

%%%%%
 \bibitem{Vulic} Vuli\'c M., Kolter R., 2001. Evolutionary cheating in Escherichia coli stationary phase cultures. {\it Genetics}, 158(2), pp.519-526.%\different cheater bacteria

%%%%%
\bibitem{Wakano} Wakano, J.Y., Nowak, M.A. and Hauert, C., 2009. Spatial dynamics of ecological public goods. {\it Proc Natl Acad Sci}, 106(19), pp.7910-7914.
%spatial public goods

%%%%%
\bibitem{watts2006} 
Watts, P.C., Buley, K.R., Sanderson, S., Boardman, W., Ciofi, C. and Gibson, R., 2006. Parthenogenesis in Komodo dragons. {\it Nature}, 444(7122), pp.1021-1022.






 
\end{thebibliography}
\end{document}